\documentclass[%
 twocolumn,
superscriptaddress,
nobibnotes,
 amsmath,amssymb,
prapplied,
floatfix,
citeautoscript
]{revtex4-1}

\usepackage[utf8x]{inputenc}

\usepackage{graphicx}
\usepackage{dcolumn}
\usepackage{bm}

\usepackage{tablefootnote}
\usepackage{threeparttable}

\usepackage{hyperref}
\hypersetup{
    colorlinks=true,
    linkcolor=blue,
    filecolor=magenta,      
    urlcolor=cyan,
}

\newcommand{\beginsupplement}{%
        \setcounter{table}{0}
        \renewcommand{\thetable}{SM\arabic{table}}%
        \setcounter{figure}{0}
        \renewcommand{\thefigure}{SM\arabic{figure}}%
        \setcounter{equation}{0}
        \renewcommand{\theequation}{SM\arabic{equation}}%
        \setcounter{section}{0}
        \renewcommand{\thesection}{SM\arabic{section}}%
     }

\usepackage{xcolor}
\usepackage[draft]{changes}

\definecolor{AKcolor}{rgb}{0.7,0.1,0.3}

\definechangesauthor[color=blue]{common}
\definechangesauthor[color=red, name={Paulo Santos}]{PVS}
\definechangesauthor[color=AKcolor, name={Alexander Kuznetsov}]{AK}
\setauthormarkup{}

\newcommand{\lBAW}{{\lambda_{\mathrm{SAW}}}}		
\newcommand{\fBAW}{f_\mathrm{AW}}			          

\begin{document}

\preprint{Electrically driven polariton optomechanics ...}

\title{Electrically driven exciton-polariton optomechanics at super high frequencies}


\author{Alexander S. Kuznetsov*}
\affiliation{Paul-Drude-Institut f{\"u}r Festk{\"o}rperelektronik, Leibniz-Institut im Forschungsverbund Berlin e.~V., Hausvogteiplatz 5-7, 10117 Berlin, Germany }

\author{Diego H. O. Machado}
\affiliation{Paul-Drude-Institut f{\"u}r Festk{\"o}rperelektronik, Leibniz-Institut im Forschungsverbund Berlin e.~V., Hausvogteiplatz 5-7, 10117 Berlin, Germany }
\affiliation{UNESP, S{\~a}o Paulo State University, Department of Physics, 
Av. Eng. Luiz Edmundo C. Coube 14-01, 17033-360, Bauru, SP, Brazil.}

\author{Klaus Biermann}
\affiliation{Paul-Drude-Institut f{\"u}r Festk{\"o}rperelektronik, Leibniz-Institut im Forschungsverbund Berlin e.~V., Hausvogteiplatz 5-7, 10117 Berlin, Germany }

\author{Paulo V. Santos}
\affiliation{Paul-Drude-Institut f{\"u}r Festk{\"o}rperelektronik, Leibniz-Institut im Forschungsverbund Berlin e.~V., Hausvogteiplatz 5-7, 10117 Berlin, Germany }
 \email{santos@pdi-berlin.de}

\date{\today}

\begin{abstract}
Polaritons enable the resonant coupling of excitons and photons to vibrations in the application-relevant super high frequency (SHF, 3-30 GHz) domain. We introduce a novel platform for coherent optomechanics based on the coupling of exciton-polaritons and electrically driven SHF longitudinal acoustic phonons confined within the spacer region of a planar Bragg microcavity.An intrinsic property of the microcavity platform is the back-feeding of phonons via reflections at the sample boundaries, which enables frequency \(\times\) quality factors products exceeding \(10^{14}\)~Hz as well as huge modulation amplitudes of the optical transition energies (up to 8 meV). We show that the modulation is dominated by the phonon-induced energy shifts of the excitonic polariton component, thus leading to an oscillatory transition between the regimes of weak and strong light-matter coupling. These results open the way for polariton-based coherent optomechanics in the non-adiabatic, side-band-resolved regime of coherent control.
\end{abstract}

\maketitle



\section{INTRODUCTION}

The coherent coupling between photons and mechanical vibrations (termed optomechanics~\cite{Aspelmeyer_RMP86_1391_14}) has experienced substantial theoretical and experimental advances since the initial investigations of  parametric instabilities in Fabry-Perot  interferometers ~\cite{Braginsky_PLA287_331_01} and of the coherent optical excitation of mechanical motion in MHz range~\cite{Kippenberg_PRL95_33901_05}. In particular, the demonstration of the strong optomechanical coupling in the MHz range~\cite{Groblacher_N460_724_09}, laser cooling of a microcavity (MC) to the mechanical ground  state~\cite{Chan_N478_89_11}, quantum-coherent coupling of NIR photons and MHz phonons~\cite{Verhagen_N482_63_12}, and optomechanically induced transparency~\cite{Weis_S330_1520_10} constitute important landmarks in this field. 

The strong-coupling between photons and quantum well excitons results in MC exciton-polariton (MP) quasiparticles, which inherit the photon-like low effective mass and long-range spatial coherence from the photonic component as well as the strong exciton-like nonlinearities~\cite{Weisbuch92a}. MPs are solid-state analogues to ultra-cold atoms exhibing Bose-Einstein condensation (BEC)~\cite{Kasprzak_N443_409_06}, bistability and quantum correlations in the 10-300 K temperature range~\cite{Sanvitto_NM15_1061_16}.  The strength of the polariton interaction with vibrations normally exceeds the one for photons since the photon-related radiation pressure mechanism becomes complemented by the strong deformation potential modulation of the  excitonic resonances.~\cite{PVS156} The acoustic modulation of MPs  has so far only  been demonstrated for sub-GHz monochromatic strain fields induced by electrically excited surface acoustic waves (SAWs)~\cite{PVS169} and for transient strain fields with frequencies in the several-GHz to THz range produced by short laser pulses~\cite{Berstermann_PRB86_195306_12}.  
In the latter case, one can potentially reach the non-adiabatic regime, where the phonon energy quantum exceeds both the phonon 
($\Gamma_\mathrm{a}=\hbar\omega_\mathrm{a}/Q_\mathrm{a}$) 
and polariton ($\Gamma_\mathrm{pol}=\hbar\omega_\mathrm{pol}/Q_\mathrm{pol}$)  decoherence rates,  thus leading to side bands in the optical spectrum shifted by multiples of the phonon energy.~\cite{Berstermann_PRB80_75301_09,Metcalfe_PRL105_37401_10}. In the previous expressions, $\hbar\omega_i$ and $Q_i$ denote the energy and quality factor of the polariton ($i=\mathrm{pol}$) and phonon ($i=\mathrm{a}$) resonances, respectively. 

MP optomechanics profits from the ability of planar (Al,Ga)As MCs to confine simultaneously light and phonons within the same spatial region.\cite{Trigo02a,Fainstein_PRL110_37403_13,Kyriienko_PRL112_76402_14}.
The latter relies on the approximately constant ratio between the impedances and propagation velocities for light and sound in  Al\textsubscript{x}Ga\textsubscript{1-x}As alloys with different compositions $x$~\cite{Fainstein_PRL110_37403_13}.  
As a consequence of the higher optical velocities, an (Al,Ga)As-based MC designed for near-IR photons also confines  longitudinal acoustic phonons in the form of GHz bulk acoustic waves (BAWs)~\cite{Fainstein_PRL110_37403_13}. The resonant enhancement of the optomechanical coupling in these MCs was demonstrated in Refs.~\onlinecite{PVS169,Rozas_PRB90_201302_14} and attributed to the large photoelastic coupling at the MP resonances~\cite{Jusserand_PRL115_267402_15}. 

\begin{figure*}[!thbp]
\includegraphics[width=0.9\textwidth, keepaspectratio=true]{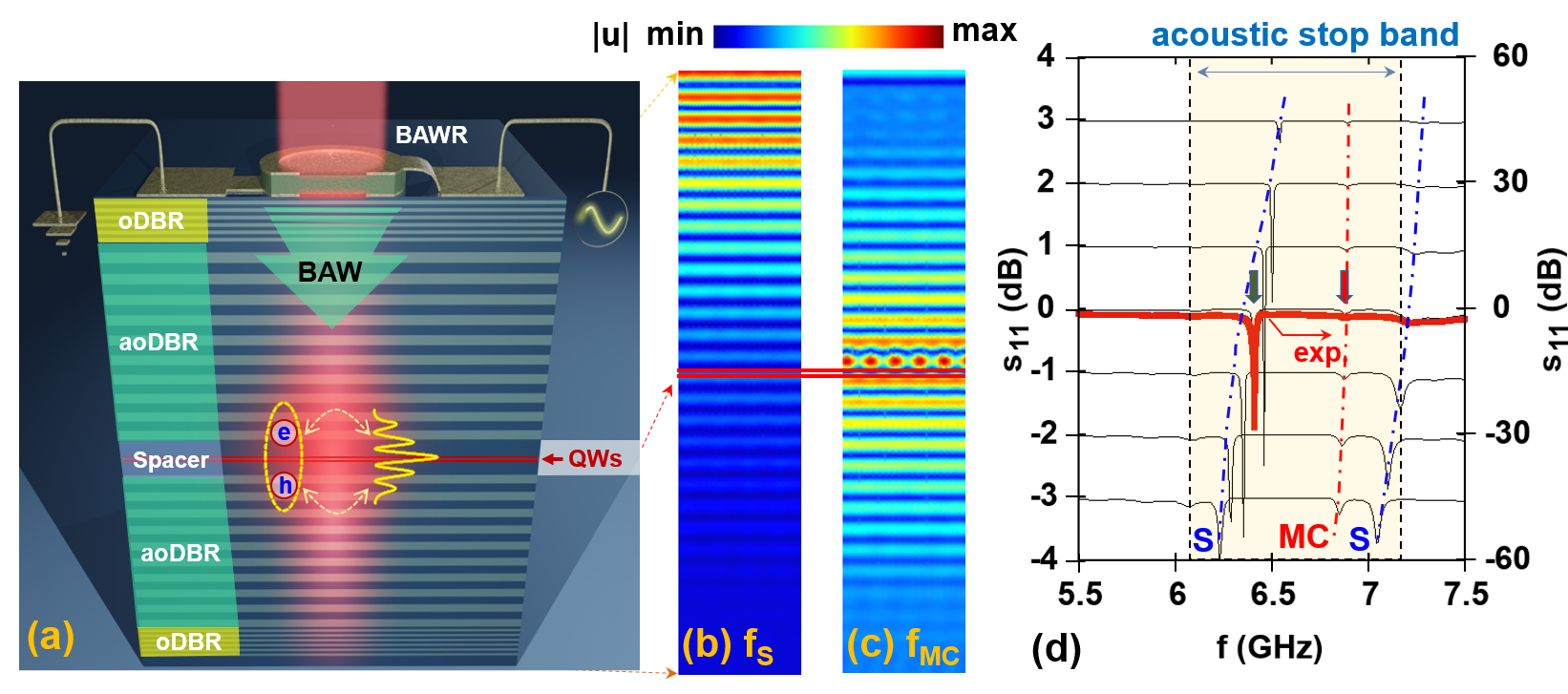}
\caption{{\bf Hybrid microcavity (MC) for polaritons and phonons.} (a) Schematic cross-section of Sample A showing the spatial distribution of photon (red-shared region) and exciton-polariton fields (yellow). oDBR and aoDBR stand for the distributed Bragg reflectors (DBRs) acting as optical and acoustic mirrors, respectively. Longitudinal bulk acoustic waves (BAWs) are generated by a ring-shaped bulk acoustic wave resonator (BAWR) driven by rf-voltage. The ring-shaped BAWR has an aperture for laser excitation  of the MC. (b)-(c) Profiles for the acoustic displacement field $|{\bf u}|$  calculated for the mode localized (b) at the sample surface (mode with f\textsubscript{S} = 6.36 GHz) and (c) at the MC spacer [f\textsubscript{MC} = 6.83 GHz, cf. thick arrows in (d)]. (d) $s_\mathrm{11}$ rf-scattering parameter of the BAWR (thick red curve) and calculated  $s_\mathrm{11}$ profiles for different thicknesses  d\textsubscript{ZnO} of the piezoelectric ZnO layer of the BAWR varying from 300 nm (bottom curve) to 180 nm (top curve) in steps of 20 nm (thin black lines). The calculations were done according to the procedure outlined in Ref.~\cite{PVS327}. The colored area in (d) indicates the spectral extent of the acoustic stopband of the MC.}
\label{Fig1}
\end{figure*}

Previous optomechanical studies in planar MCs  mostly employed phonons in the form of electrically excited, sub-GHz SAWs~\cite{PVS223} or BAWs in the GHz range stimulated either thermally or optically using short optical pulses. Here, we introduce a platform for electrically driven MP optomechanics in the SHF region (3-20~GHz) based on phonon generation and detection using high-frequency BAW resonators (BAWRs).  BAWRs enable acoustic  echo spectroscopy with a very high (over 90~dB) dynamic range~\cite{PVS327}. 
The latter is important to unveil the distribution of the acoustic field within the samples, which  results from the resonant coupling of BAWs modes confined in three coupled acoustic cavities: the main one within the MC spacer, a surface cavity between the upper and lower distributed Bragg reflectors (DBRs) and a bulk cavity formed by the front and back surfaces of the wafer. 
The coupling enables the back-feeding of BAWs into the main cavity, thus   resulting in  acoustic quality factors $Q_{a}$'s significantly higher than the ones expected from the DBR acoustic reflectivity as well as in frequency $\times$ $Q_{a}$ products exceeding $10^{14}$.
The strong acoustic field in the QW region induces a huge modulation of the MP energies, which reaches amplitudes (up to 8~meV) far exceeding the light-matter Rabi coupling. We show that the large energy modulation arises from the deformation potential coupling to the MP excitonic components, which  leads to an oscillatory transition between the regimes of weak and strong light-matter coupling at GHz frequencies.

\begin{figure}[!thbp]
\includegraphics[width=0.95\columnwidth, keepaspectratio=true]{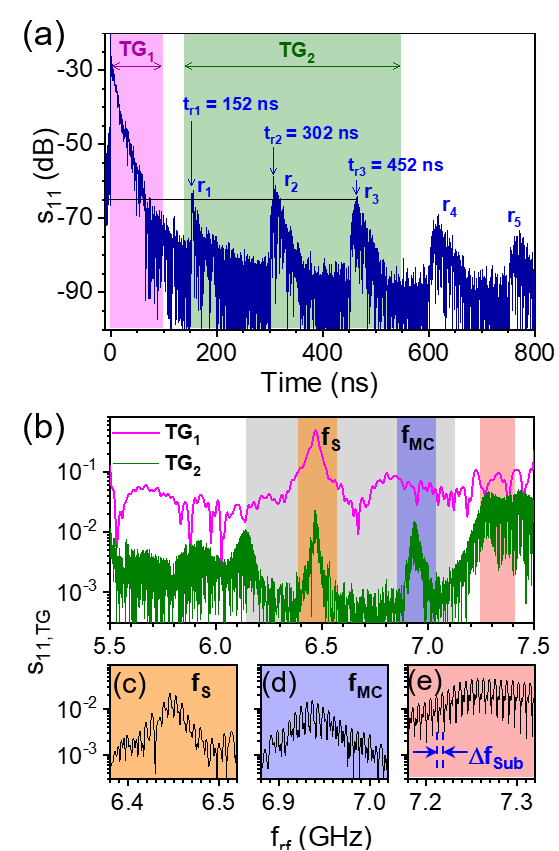}
\caption{{\bf Acousto-electric response of a  hybrid MC at 10 K}. (a) Time-dependence of rf BAWR reflection of Sample A determined from the spectral response of the $s_\mathrm{11}$ rf-scattering parameter in the 5.5-7.5 GHz spectral range. The multiple acoustic echoes r\textsubscript{i} (i = 1 to 5) result from acoustic reflections at the back surface of the substrate. (b) Spectral dependence of the echoes within the time ranges TG\textsubscript{1} (0-100 ns, pink) and TG\textsubscript{2} (150-550 ns, green, encompassing three acoustic echoes) defined in (a). The grey area is the stopband range. Two acoustic modes can be seen within the stopband for the TG\textsubscript{2} (see text for discussion). (c)-(e) Close-ups of the green curve in (b) highlighting the frequency comb with spacing $\Delta f_\mathrm{Sub} = 6.4$~MHz.}\label{Fig2}
\end{figure}

\section{RESULTS AND DISCUSSION}

The studies were carried out on hybrid (Al,Ga)As MCs designed to simultaneous confine polaritons and phonons within the spacer region containing two (In,Ga)As quantum wells (see Methods). We first present results for a MC designed for optical  and phonon wavelengths $\lambda_o=850/n_\mathrm{GaAs}$~nm and $\lambda_a=3\lambda_o$, respectively, where $n_\mathrm{GaAs}$ is the GaAs refractive index  [Sample A, Fig.~\ref{Fig1}(a)]. 
This sample simultaneously confines  in the MC spacer photons with a wavelength of 850~nm and phonons with a frequency of approximately 7~GHz.
The acousto-optical distributed Bragg reflector (aoDBR) acts as a first-order  and third-order grating for phonons and photons, respectively. The optical DBR (oBDR) provides the additional optical confinement required to achieve the strong coupling regime. The BAWs were electrically excited by a BAWR with ring-shaped electrodes. This design helps to concentrate the acoustic field at the aperture center and provides optical access to the spacer region (cf. Supplementary Material, Sec.~\ref{Lateral field distribution in ring-shaped BAWR}).

The thick red line in Fig.~\ref{Fig1}(d) displays the measured $s_{11}$ rf-scattering for the BAWR (corresponding to the rf power reflection coefficient) over a frequency range covering the acoustic stopband region of the aoDBR. Two resonances (denoted as MC and S) are found within the stopband range. The thin curves are the finite-element calculations of the  $s_\mathrm{11}$ response for varying thicknesses $d_\mathrm{ZnO}$ of the ZnO layer, which accurately reproduce the measured response for  nominal thickness $d_\mathrm{ZnO} = 240$~nm (thick line) \cite{getdp-ieee1998,PVS327}. While the MC-mode is essentially insensitive to $d_\mathrm{ZnO}$, the S-mode shifts towards lower frequencies with decreasing $d_\mathrm{ZnO}$. This behavior arises from the fact that the S-mode is confined between the upper aoDBR and the BAWR surface,  while the MC-mode is concentrated in-between aoDBRs, as illustrated by the calculated mode profiles of Figs.~\ref{Fig1}(b) and \ref{Fig1}(c), respectively.

The time response obtained via a Fourier transformation of the $s_\mathrm{11}$ over the frequency range of the acoustic stopband yields additional information about the interplay between the acoustic modes [cf.~Fig.~\ref{Fig2}(a)]. The time trace is characterized by an exponentially decaying signal at short times [$<100$~ns, denoted as TG\textsubscript{1} in Fig.~\ref{Fig2}(a)] followed by a series of echoes delayed by t\textsubscript{rt}= 151 \ensuremath{\pm} 1 ns (region TG\textsubscript{2}). The echoes are associated with round-trips of BAWs reflected at the backside of the double-polished GaAs substrate. Indeed, by taking the LA phonon velocity in GaAs $v_\textsubscript{LA}= 4.7~\mu$m/ns and the nominal substrate thickness $d_\textsubscript{sub} = 350 \pm 20~\mu$m, one obtains a round-trip time delay $2d\textsubscript{sub}/v\textsubscript{LA} = 149 \ensuremath{\pm} 15$~ns, matching the value of t\textsubscript{rt}. Although only 5 reflections are shown in Fig.~\ref{Fig2}(a), up to 9 echoes could be detected thus yielding a BAW  lifetime exceeding $0.3~{\mu}$s. 

The spectral contributions of the individual acoustic cavities can be identified by an inverse Fourier transformation of the time trace in  Fig.~\ref{Fig2}(a) within the TG\textsubscript{1} and TG\textsubscript{2} delay regions [cf. spectra $s_\mathrm{11,TG}$ displayed Fig.~\ref{Fig2}(b)]. The acoustic response at short times (TG\textsubscript{1} range) is dominated by a single strong resonance at $f_\mathrm{S} = 6.46$~GHz corresponding to the surface cavity resonance mode  of  Fig.~\ref{Fig1}(b). The confinement near the surface by the upper aoDBR also accounts for the long decay time of the echos in Fig.~\ref{Fig2}(a), which far exceeds the short transit time ($\sim 1/f_S=0.15$~ns) across the BAWR. The spectrum of the TG\textsubscript{2} time range shows two peaks (denoted as $f_{S}$ and $f_{MC}$) located on a wide background of low values of the  $s_\mathrm{11,TG}$ response (gray-shaded background). The latter is attributed to the acoustic stopband of the aoDBRs, which can be detected due to the high dynamic range and time resolution (or, equivalently, wide frequency response) of the BAWRs. The two peaks at $f_{S}=6.46$~GHz and $f_{MC} = 6.94$~GHz are attributed to acoustic modes of the surface (S) and the main (MC) cavities, respectively. The absence of the mode at $f_{MC}$ in the spectrum for the $TG_{1}$ is likely due to the large background induced by incomplete suppression of the electromagnetic contribution at short echo delays. 

A closer examination of the frequency response of the TG$_\textsubscript{2}$ range reveals a frequency comb with the free spectral range (FSR)  $\Delta f_\mathrm{Sub} = 1/t_\mathrm{r1} =6.4$~MHz [cf. Figs.~\ref{Fig2}(c)-(e)]. 
The  transmission through a Bragg resonator approaches unity close to the resonance frequency. The resonance modes of the surface and main acoustic cavities thus propagate through the whole structure. The constructive interference of BAWs after multiple round-trips results in the frequency comb -- phonon backfeeding. The  quality factors of the comb resonances reach values of $Q\textsubscript{a} \ge 2800$ at 6.937~GHz, which are considerably larger than the bare quality factor ($Q\textsubscript{a,MC} = 172$) of the main cavity calculated from the reflectivity of the aoDBRs. We show in Sec.~\ref{Frequency response of bulk acoustic resonators} that due to the phonon back-feeding the effective acoustic quality factor becomes $Q_\mathrm{a} = Q_\mathrm{a,Sub}  \left( 1 + \frac{d_s}{d_\mathrm{Sub}} Q_\mathrm{a,MC} \right)$, where $Q_\mathrm{a,Sub}$ is the quality factor of the bare substrate and $d_s/d_\mathrm{Sub}$ is the thickness ratio between the MC spacer and the substrate. The enhancement of the quality factor leads to a very large $Q_{a} \times f\sim2*10^{13}$~Hz product for the comb resonances around 6.94 GHz.

We now turn our attention to the interaction between BAWs and MPs. The MPs result from the strong coupling between the MC photons and excitons in the two InGaAs QWs inserted into the MC spacer of Sample A (see Methods and Sec.~\ref{Temperature dependence of the opto-electronic resonances}). 

\begin{figure}[!thbp]
\includegraphics[width=1\columnwidth, keepaspectratio=true]{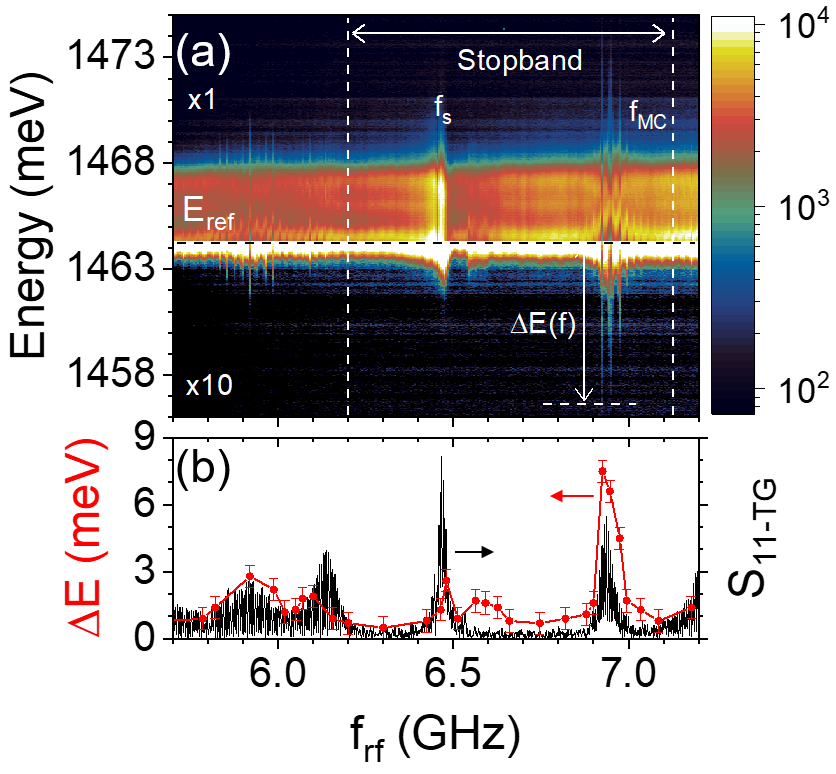}
\caption{{\bf Polariton-phonon interaction in a hybrid microcavity}. 
(a) Spectral dependence of the photoluminescence on the rf-frequency for Sample A recorded at 10~K and a nominal rf-power of 24~dBm. The acoustic stopband limits are indicated by the vertical dashed lines.  (b) Comparison of the time-gated S$_{11}$ parameter of a similar device with  the acoustically-induced energy modulation amplitude of the comb resonances $\Delta$E(f), calculated with respect to the E$_{ref}$ = 1464.3 meV (indicated in the panel-a with the horizontal black dashed line).
}\label{Fig3}
\end{figure}

\begin{figure*}[!thbp]
\centering\includegraphics[width=0.9\textwidth, keepaspectratio=true]{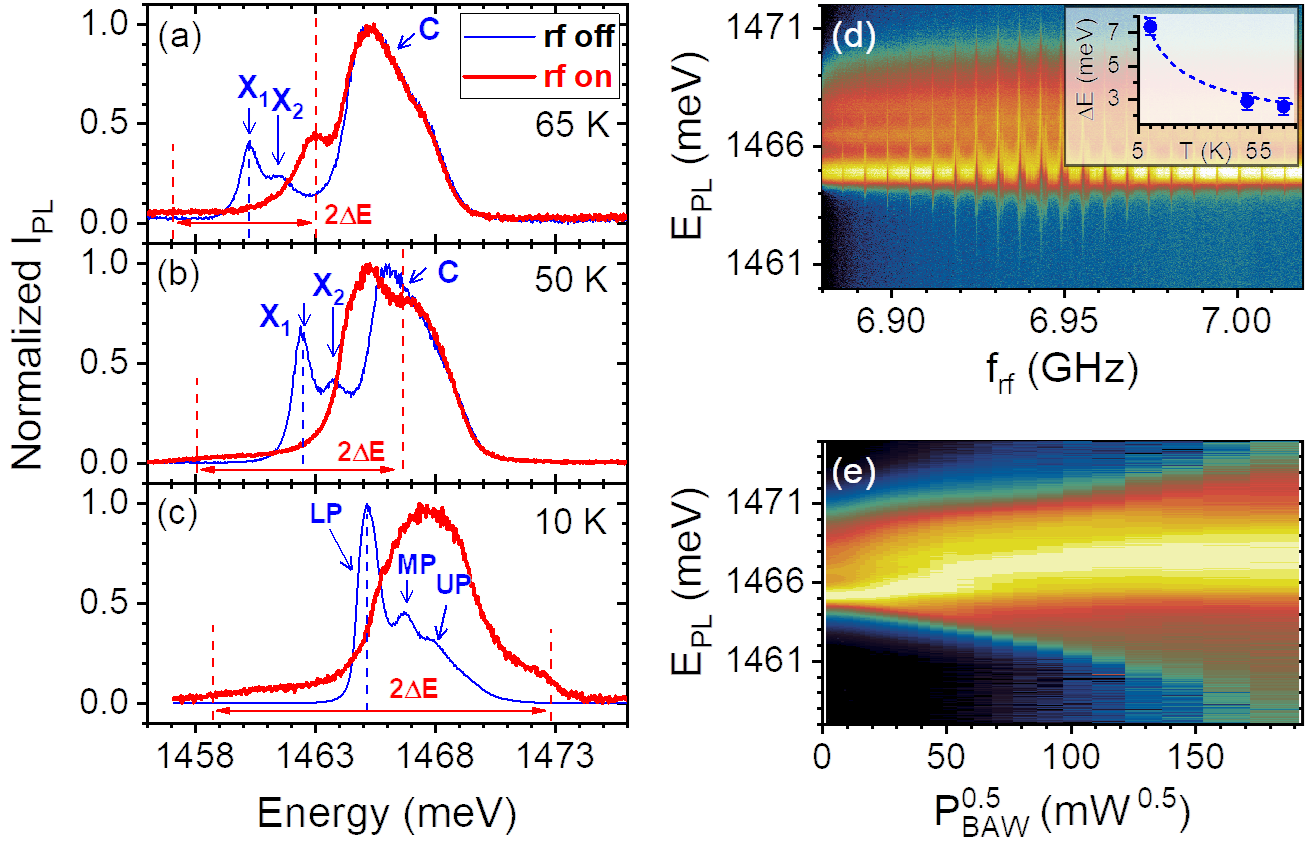}
\caption{{\bf Modulation of polaritons by MC acoustic modes.} Time-averaged PL spectra of Sample A recorded in the absence (thin blue lines) and presence (thick red lines) of a BAW with frequency $f\textsubscript{BAW} =6.931$~GHz recorded at (a) 65~K, (b) 50~K, and (c) 10~K. $\Delta E$ denotes the energy modulation amplitude of the excitonic levels.
(d) rf-frequency  dependence of the PL recorded  for a fixed rf-power P\textsubscript{rf} = 14 dBm applied to the BAWR at 10~K, corresponding the coupled linear power amplitude P$_{BAW}^{0.5}$ = 45 mW$^{0.5}$, showing the effects of the frequency comb with $\Delta f_\mathrm{Sub}= 6.4$~MHz. The inset displays the temperature (T) dependence of the energy modulation amplitude $\Delta E$. The dashed line in the inset is a guide to the eye. (e) Dependence of the PL recorded for a fixed rf-frequency f$_{rf}$ = 6.9312 GHz applied to the BAWR at 10~K on P$_{BAW}^{0.5}$.}
\label{Fig4}
\end{figure*}

The color map of Fig.~\ref{Fig3}(a) shows the time-integrated polariton PL as a function of the  applied rf-frequency. The acoustic modulation induces an energy modulation (of amplitude $\Delta E$) as well as changes in PL intensity at the comb frequencies. 
Figure~\ref{Fig3}(b) compares the spectral dependence of $\Delta E$ with the electrical response $S_\mathrm{11,TG}$. The PL comb is observed outside the DBR stopband as well as at resonances $f_S$ and $f_{MC}$ within the stopband. The spectral dependence of the energy modulation amplitude $\Delta E$ correlates very well with $S_\mathrm{11,TG}$. In particular, $\Delta E$ peaks at the main cavity resonance $f_{MC}$, where the acoustic field becomes confined within the MC spacer, thus increasing the interaction with QW excitons. Figure~\ref{Fig4}(d) displays spectral PL profiles around this frequency. The sharp PL comb lines have an effective quality factor comparable to the one determined from the $S_\mathrm{11,TG}$ curves of Fig.~\ref{Fig2}(b). The PL energy modulation reaches amplitudes up to $\Delta E = 8$~meV, which exceed several times the Rabi-splitting energy of approx. $\Omega_\mathrm{Rabi}$ =  2~meV (see Sec.~\ref{Temperature dependence of the opto-electronic resonances}). Figure~\ref{Fig4}(e) shows that $\Delta E$ can be continuously tuned by changing the rf-power applied to the BAWR.

In order to unveil the mechanisms responsible for the phonon-induced modulation of the MP energies, we  compare in Figs.~\ref{Fig4}(a)-(c) time-averaged PL spectra of the MC recorded at different temperatures in the absence of acoustic excitation (thin blue lines) with the ones acquired under the  excitation at a comb frequency in the $f_\mathrm{MC}$ range (thick red lines). The  InGaAs QWs in Sample A are separated by a narrow (5~nm-thick) GaAs barrier, which tunnel-couples their excitonic states to produce bonding ($X_1$) and anti-bonding resonances ($X_2$). At temperatures above 50~K, these states are red-shifted and only weakly couple to the cavity resonance ($C$), giving rise to  the three PL peaks in Figs.~\ref{Fig4}(a) and (b) [further details in Sec.~\ref{Temperature dependence of the opto-electronic resonances}].

When the BAW is turned-on (red curves in Figs.~\ref{Fig4}(a) and \ref{Fig4}(b)), the sinusoidal modulation of the excitonic energies with amplitude $\Delta E$ leads to two shoulders in the time-averaged PL spectra shifted by $\pm\Delta E$ with respect to the unperturbed excitonic energy (dotted vertical blue lines that correspond to the unperturbed X$_1$ excitonic resonances)~\cite{PVS107}. The red-shifted shoulder is much weaker than the blueshifted one (dotted vertical red lines) since the large detuning between the excitonic and photonic resonances reduces their coupling to the photonic mode. More importantly, the photonic mode ($C$  at 1465 meV) remains essentially unperturbed, thus indicating that the energy modulation is dominated by the strain-induced modulation of the excitonic resonances.

At lower temperatures, the excitonic lines blueshift and strongly couple to the photonic mode to form the lower (LP), middle (MP), and upper (UP) polariton states indicated in Fig.~\ref{Fig4}(c) [cf. Sec.~\ref{Temperature dependence of the opto-electronic resonances}].
The BAW-induced energy modulation in Fig.~\ref{Fig4}(c) is sufficiently large to blueshift the excitonic levels beyond the regime of the strong coupling, thus leading to the appearance of a shoulder at a huge blueshift of approximately $\Delta E =8$~meV [dotted vertical line in Fig.~\ref{Fig4}(c)]. 

The modulation of the exciton energy is attributed to the deformation potential mechanism, which yields  $\Delta E = (a_h) \eta_s u_{zz,0}$. Here, $\eta_s u_{zz,0}$ is the amplitude of the strain field at the QW position, which is factor $\eta_s\sim0.8$ smaller than the amplitude $u_{zz,0}$ of the strain field in the MC spacer (see Sec.~\ref{Field distribution in hybrid microcavities}), and  $a_h\approx 10~$eV is the hydrostatic deformation potential for electron-hole transitions \cite{LB17a}.   
From the $\Delta E$ value in Fig.~\ref{Fig4}(c) we obtain  $\eta_s u_\mathrm{zz,0}=8\times10^{-4}$. Since the amplitude of the phonon displacement field $u_{z,0}{\hat z}$ is given by $u_{z,0}=({\lambda_a}/{2\pi}) u_\mathrm{zz,0}$, the effective opto-mechanical coupling is estimated to be 
$g_\mathrm{eff}=\Delta E/u_{z,0} = (\eta_s 2 \pi/\lambda_a)a_h = 18$~THz/nm.
 
Another  interesting observation in Fig.~\ref{Fig4} is the strong increase of  $\Delta E$ with decreasing temperature (T) for a fixed rf-excitation [cf. symbols in lower inset of Fig.~\ref{Fig4}(d)]. In a recent study, we showed that the acoustic propagation losses increase with the temperature in similar GaAs substrates \cite{PVS323}. The large increase in $\Delta E$ with decreasing temperature is thus attributed to the reduction of acoustic losses. 
Indeed, according to Eq.~7 of Ref.~\cite{PVS327}, the effective acoustic absorption coefficient $\alpha_\mathrm{eff}$ doubles when the temperature increases from 10~K to 65~K. Since the amplitude of the acoustic field scales with $1/\alpha_\mathrm{eff}$, the increased absorption correlates well with the two times reduction of $\Delta E$ (from 7.5 meV to 3.5 meV) from 10 K to 65 K, cf.~inset of Fig.~\ref{Fig4}(c).

Finally, the MC-based optomechanical platform can be straightforwardly scaled to higher phonon frequencies. Here, one takes advantage of the fact that effective phonon absortion losses $\alpha_\mathrm{eff}$ at low temperatures remain approximately constant in the 3-30~GHz frequency range, thus resulting in an increase of the quality factor with frequency~\cite{PVS327}. In the following, we present results for a polariton MC with  aoDBRs designed as first-order reflectors for both photons and phonons with the same wavelength $\lambda_o=\lambda_a$  (Sample B, cf. Methods), resulting in  acoustic resonance frequencies close to 20~GHz~\cite{PVS327}. The energy quanta of these phonons of 80 µeV is several times larger than the linewidth of the BECs in similar samples~\cite{PVS255,PVS333}.  Thus, the demonstration of energy modulation amplitudes  far exceeding the polariton linewidths at these frequencies constitutes an important milestone to reach the non-adiabatic,  sideband-resolved modulation regime.

\begin{figure}[t]
\setlength{\belowcaptionskip}{-11pt}
\includegraphics[width=0.95\columnwidth, keepaspectratio=true]{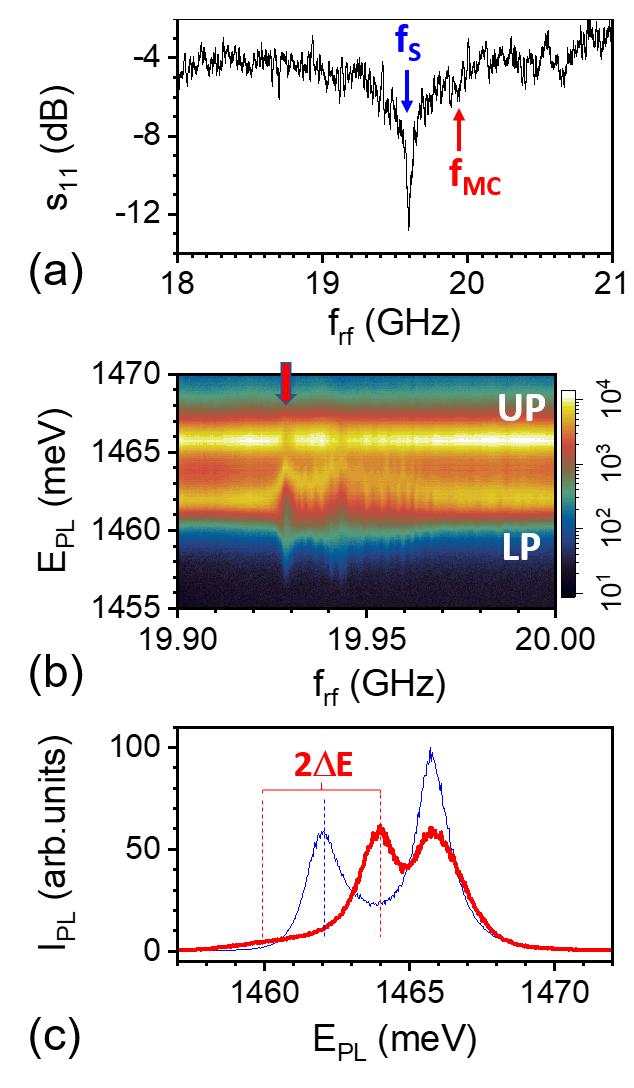}
\caption{ {\bf Optomechanical response of a hybrid MC for 20~GHz BAWs}. (a) $s_\mathrm{11}$ 
rf-scattering parameter for Sample B showing the surface ($f\mathrm{s} = 19.6$~GHz) and MC ($f_\mathrm{MC} \approx 20$~GHz)  acoustic resonances. (b) The rf-frequency  dependence of the PL recorded  for a fixed rf-power P\textsubscript{rf} = 24~dBm applied to the BAWR displaying the effect of the BAW on the PL emission. (c) Time-averaged PL spectra recorded at 10~K in the absence (thin blue line) and presence (thick red line) of a BAW with frequency $f\textsubscript{BAW} =$19.926~GHz [cf. red arrow in (b)]. LP and UP denote the exciton-like and photon-like polariton brunches, respectively. All measurements were carried out at 10 K.}
\label{Fig5}
\end{figure}

The electrical response of this sample [cf.~Fig.~\ref{Fig5}(a)] displays the signatures of the surface ($f_S$) and MC mode ($f_{MC}$) at a frequency approximately three times higher than for Sample A, cf. Fig.~\ref{Fig1}(d). A detailed analysis of the $s_{11}$ response shows a frequency comb within the $f_\mathrm{MC}$ frequency range consisting of resonances with an effective quality factor $Q_a=6800$ and a huge product $Q_{a} \times f\sim 1.3*10^{14}$~Hz [cf. Sec.~\ref{Frequency response of bulk acoustic resonators}]. The ratio between the effective quality factors of samples B and A of 2.4 compares reasonably well with the ratio of 2.8 between their resonance frequencies.

Driving the BAWR within the $f_\mathrm{MC}$ frequency range (19.92--19.97~GHz) induces a comb of resonances in the PL spectrum, as illustrated in  Fig.~\ref{Fig5}(b), which  modulates the lower-polariton energies with an amplitude $\Delta E$ up to 2~meV [cf.~Fig.~\ref{Fig5}(c)], thus yielding strain amplitudes $u_{zz}\approx 1\times 10^{-4}$ and 
$g_\mathrm{eff}=50$~THz/nm 
at 20~GHz. 
The weaker modulation of the UP-polariton branch in Sample B is attributed to its mostly photonic character, cf. Fig.~\ref{FigS6}.

\section{CONCLUSIONS}

In conclusion, we have demonstrated a novel platform for electrically driven GHz exciton-polariton optomechanics in the SHF range based on the coupling between MPs and electrically generated BAWs confined in a planar MC. This platform profits, on the one hand, from the long effective lifetimes of phonons confined in the MC, where they are less susceptible to the degradation of the quality factor due to the surface quality, thus yielding $Q_{a}\times f$ products in the SHF range exceeding $10^{14}$~Hz, which  are comparable to the highest values reported for much lower vibration frequencies ~\cite{Aspelmeyer_RMP86_1391_14,Hamoumi_PRL120_223601_18,PVS327}.  
On the other hand, it exploits the high sensitivity of the excitonic resonances to confined strain fields, which enables 
the modulation amplitudes of the MPs energies far exceeding the light-matter coupling strength as well as effective optomechanical couplings in the $g_\mathrm{eff}\sim 20$~THz/nm range. 
As a prospect, the results open the way to resonant optomechanics in the non-adiabatic, side-band limited regime using polariton condensates with spectral linewidths considerably smaller ($<10~\mu$eV)~\cite{PVS255}) than the ones for the sub-condensation regime reported here. In  particular, polariton linewidths below the inverse phonon frequency will enable the study of interesting phenomena such as mechanical self-oscillations and phonon lasing. Finally, the recent developments in MC structures have demonstrated the feasibility of zero-dimensional confinement of polaritons~\cite{Schneider_RPP_16503_17,PVS312} and phonons~\cite{Anguiano_PRL118_263901_17} in structured MCs. Electrically driven, high frequency BAWs in these structures thus  provide access to the single-phonon regime at temperatures (of $\sim 1$~K) substantially larger than for sub-GHz vibrations.

\section*{Methods}

The hybrid planar (Al,Ga)As MCs for the polariton-phonon confinement [cf.~Fig.~\ref{Fig1}(a)] were grown on GaAs (001) substrates  by molecular beam epitaxy. 
In Sample A (cf.~Fig.~\ref{Fig1}), the outer optical DBR (oDBR) stacks  consist of six pairs of \ensuremath{\lambda}/4-thick GaAs/Al\textsubscript{0.85}Ga\textsubscript{0.15}As layers and provide optical confinement. Here, $\lambda=\lambda_{0}/n_{i}$ is the optical resonance wavelength in medium i with refractive index $n_{i}$ for a free space wavelength of \ensuremath{\lambda}\textsubscript{0} = 850~nm. The inner acoustic DBR (aoDBR) stacks ensure acoustic as well as optical confinement. It consist of 10 GaAs/Al\textsubscript{0.85}Ga\textsubscript{0.15}As layer pairs with thickness per layer of 3\ensuremath{\lambda}/4, thus acting as a first order acoustic DBR for longitudinal phonons with a center wavelength $\lambda_{a}=250$~nm and as a third order optical grating for a free space optical wavelength \ensuremath{\lambda}\textsubscript{0}. The spacer region of Sample A concurrently acts as an optical $5\lambda/2$  and as an acoustic $2\lambda_a/2$ MC spacer. This region embeds two 15~nm-thick In\textsubscript{0.04}Ga\textsubscript{0.96}As quantum wells (QW) separated by a 5 nm-thick GaAs barrier. The latter are positioned at a depth corresponding to an antinode of both the optical and the acoustic strain fields inside the MC spacer (see further details in Sec.~\ref{Field distribution in hybrid microcavities} of the Supplementary Material, SM). Transfer matrix simulations  were used to estimate an optical quality factor Q\textsubscript{o} 
= 5600 and a Rabi splitting $\Omega_\mathrm{Rabi} = 4.3$~meV at 10~K. The measured $\Omega_\mathrm{Rabi}$ is about 2 meV (cf.~Sec.~\ref{Temperature dependence of the opto-electronic resonances}). Simulations for the acoustic field yield a bare (i.e., neglecting BAW reflections at sample borders, see below for details) acoustic quality factor Q\textsubscript{a} = 172 at 6.9 GHz at 10 K.

Sample B consists of a 2$\lambda/2$ wide spacer including a single, 15~nm thick In\textsubscript{0.04}Ga\textsubscript{0.96}As QW. The spacer is sandwiched between DBRs with $\lambda/4$ layers, which act as first order reflectors for 850 nm photons and 20 GHz phonons. The anti-nodes of the photon and pnonon fields overlap at the QW position. The upper and lower RBRs consists of 20 and 36 pairs of \ensuremath{\lambda}/4-thick GaAs/Al\textsubscript{0.85}Ga\textsubscript{0.15}As, respectively. The sample is in the strong-coupling regime at 10~K with Rabi-splitting energy $\Omega_\mathrm{Rabi} = 3$~meV, cf. Fig. \ref{FigS6}.

The BAWs were excited by bulk acoustic wave resonators (BAWRs)~\cite{Lakin_ITUFFC52_707_05,PVS327} fabricated on the sample surface, as illustrated schematically in Fig.~\ref{Fig1}(a). The active region of the BAWRs on Sample A (Sample B) consists of a nominally 260~nm-thick (70~nm for Sample B) 
textured  ZnO film sputtered with the hexagonal c-axis oriented perpendicular to the MC surface (these thickness were chosen according to the studies in Ref. [\onlinecite{PVS327}]). The piezoelectric film is sandwiched between two 50~nm thick metal contacts. A special feature of the BAWR design is the ring-shape geometry with apertures in the bottom and top contacts for optical access to the underlying MC (cf. Sec.~\ref{Lateral field distribution in ring-shaped BAWR}). The piezoelectrically active area is thus defined by the overlap region of the top and bottom electrodes. The electrical response of the BAWRs was measured using a vector network analyzer with time-gating capabilities (cf.~Sec.~\ref{Frequency response of bulk acoustic resonators}).

The spatially-resolved photoluminescence (PL) measurements were carried out at 10~K in a cryogenic (liquid He) probe station. Optical excitation was provided by a pulsed diode laser emitting at 635 nm with fluences between 10 and 200 µW focused onto a  10×2 µm$^2$ spot. The PL was spectrally resolved using a single grating spectrometer with resolution of 100 µeV and detected  by a liquid-nitrogen-cooled CCD detectors. The PL measurements under acoustic excitation were carried out using the  modulation technique described in Sec.~\ref{Compensation for thermal effects in the spectroscopic measurements} \cite{PVS282}, which compensates for eventual thermal effects arising from sample heating during the rf excitation. The typical PL acquisition times were  between 1 and 30~s per spectrum.

{\noindent {\it Authors contributions: }} 
A.S.K. has participated in the inception of the idea, proposed the design of the transducer aperture-electrodes, carried out all optical and electrical measurements and analyzed data. 
D.H.O.M. has fabricated acoustic devices. 
K.B. has designed (using optical and acoustic transfer matrix simulations) and fabricated the MC sample, and critically reviewed the manuscript. 
P.V.S has proposed the idea, provided finite element method acoustic simulations and participated in critical discussions.
A.S.K. and P.V.S. have equally contributed to the analysis of the results as well as to the preparation of the manuscript with the input from all co-authors.

{\noindent {\it Acknowledgements: }} 
We thank Dr. Timur Flissikowski for discussions and for a critical review of the manuscript. We also acknowledge the technical support from R. Baumann, S. Rauwerdink, and A. Tahraoui in the sample fabrication process.  We acknowledge financial support from the German DFG (grant 359162958), the QuantERA grant Interpol (EU-BMBF (Germany) grant nr. 13N14783), and FAPESP (Brazil, grant 2017/24311-6).

{\noindent {\it Competing interests: }} 
Authors declare no competing interests.

\IfFileExists{x:/sawoptik_databases/jabref/literature.bib}
{
\def\litdir{}
}
{\IfFileExists{x:/sawoptik_databases/jabref/literature.bib}
{   \def\litdir{x:/sawoptik_databases/jabref} }
{	\def\litdir{c:/myfiles/jabref} }
}

\bibliography{literature,mypapers}

\begin{thebibliography}{34}%
\makeatletter
\providecommand \@ifxundefined [1]{%
 \@ifx{#1\undefined}
}%
\providecommand \@ifnum [1]{%
 \ifnum #1\expandafter \@firstoftwo
 \else \expandafter \@secondoftwo
 \fi
}%
\providecommand \@ifx [1]{%
 \ifx #1\expandafter \@firstoftwo
 \else \expandafter \@secondoftwo
 \fi
}%
\providecommand \natexlab [1]{#1}%
\providecommand \enquote  [1]{``#1''}%
\providecommand \bibnamefont  [1]{#1}%
\providecommand \bibfnamefont [1]{#1}%
\providecommand \citenamefont [1]{#1}%
\providecommand \href@noop [0]{\@secondoftwo}%
\providecommand \href [0]{\begingroup \@sanitize@url \@href}%
\providecommand \@href[1]{\@@startlink{#1}\@@href}%
\providecommand \@@href[1]{\endgroup#1\@@endlink}%
\providecommand \@sanitize@url [0]{\catcode `\\12\catcode `\$12\catcode
  `\&12\catcode `\#12\catcode `\^12\catcode `\_12\catcode `\%12\relax}%
\providecommand \@@startlink[1]{}%
\providecommand \@@endlink[0]{}%
\providecommand \url  [0]{\begingroup\@sanitize@url \@url }%
\providecommand \@url [1]{\endgroup\@href {#1}{\urlprefix }}%
\providecommand \urlprefix  [0]{URL }%
\providecommand \Eprint [0]{\href }%
\providecommand \doibase [0]{http://dx.doi.org/}%
\providecommand \selectlanguage [0]{\@gobble}%
\providecommand \bibinfo  [0]{\@secondoftwo}%
\providecommand \bibfield  [0]{\@secondoftwo}%
\providecommand \translation [1]{[#1]}%
\providecommand \BibitemOpen [0]{}%
\providecommand \bibitemStop [0]{}%
\providecommand \bibitemNoStop [0]{.\EOS\space}%
\providecommand \EOS [0]{\spacefactor3000\relax}%
\providecommand \BibitemShut  [1]{\csname bibitem#1\endcsname}%
\let\auto@bib@innerbib\@empty
\bibitem [{\citenamefont {Aspelmeyer}\ \emph {et~al.}(2014)\citenamefont
  {Aspelmeyer}, \citenamefont {Kippenberg},\ and\ \citenamefont
  {Marquardt}}]{Aspelmeyer_RMP86_1391_14}%
  \BibitemOpen
  \bibfield  {author} {\bibinfo {author} {\bibfnamefont {M.}~\bibnamefont
  {Aspelmeyer}}, \bibinfo {author} {\bibfnamefont {T.~J.}\ \bibnamefont
  {Kippenberg}}, \ and\ \bibinfo {author} {\bibfnamefont {F.}~\bibnamefont
  {Marquardt}},\ }\href {\doibase 10.1103/RevModPhys.86.1391} {\bibfield
  {journal} {\bibinfo  {journal} {Rev. Mod. Phys.}\ }\textbf {\bibinfo {volume}
  {86}},\ \bibinfo {pages} {1391} (\bibinfo {year} {2014})}\BibitemShut
  {NoStop}%
\bibitem [{\citenamefont {{Braginsky}}\ \emph {et~al.}(2001)\citenamefont
  {{Braginsky}}, \citenamefont {{Strigin}},\ and\ \citenamefont
  {{Vyatchanin}}}]{Braginsky_PLA287_331_01}%
  \BibitemOpen
  \bibfield  {author} {\bibinfo {author} {\bibfnamefont {V.~B.}\ \bibnamefont
  {{Braginsky}}}, \bibinfo {author} {\bibfnamefont {S.~E.}\ \bibnamefont
  {{Strigin}}}, \ and\ \bibinfo {author} {\bibfnamefont {S.~P.}\ \bibnamefont
  {{Vyatchanin}}},\ }\href {\doibase 10.1016/S0375-9601(01)00510-2} {\bibfield
  {journal} {\bibinfo  {journal} {Physics Letters A}\ }\textbf {\bibinfo
  {volume} {287}},\ \bibinfo {pages} {331} (\bibinfo {year} {2001})},\ \Eprint
  {http://arxiv.org/abs/gr-qc/0107079} {gr-qc/0107079} \BibitemShut {NoStop}%
\bibitem [{\citenamefont {Kippenberg}\ \emph {et~al.}(2005)\citenamefont
  {Kippenberg}, \citenamefont {Rokhsari}, \citenamefont {Carmon}, \citenamefont
  {Scherer},\ and\ \citenamefont {Vahala}}]{Kippenberg_PRL95_33901_05}%
  \BibitemOpen
  \bibfield  {author} {\bibinfo {author} {\bibfnamefont {T.~J.}\ \bibnamefont
  {Kippenberg}}, \bibinfo {author} {\bibfnamefont {H.}~\bibnamefont
  {Rokhsari}}, \bibinfo {author} {\bibfnamefont {T.}~\bibnamefont {Carmon}},
  \bibinfo {author} {\bibfnamefont {A.}~\bibnamefont {Scherer}}, \ and\
  \bibinfo {author} {\bibfnamefont {K.~J.}\ \bibnamefont {Vahala}},\ }\href
  {\doibase 10.1103/PhysRevLett.95.033901} {\bibfield  {journal} {\bibinfo
  {journal} {Phys. Rev. Lett.}\ }\textbf {\bibinfo {volume} {95}},\ \bibinfo
  {pages} {033901} (\bibinfo {year} {2005})}\BibitemShut {NoStop}%
\bibitem [{\citenamefont {Groblacher}\ \emph {et~al.}(2009)\citenamefont
  {Groblacher}, \citenamefont {Hammerer}, \citenamefont {Vanner},\ and\
  \citenamefont {Aspelmeyer}}]{Groblacher_N460_724_09}%
  \BibitemOpen
  \bibfield  {author} {\bibinfo {author} {\bibfnamefont {S.}~\bibnamefont
  {Groblacher}}, \bibinfo {author} {\bibfnamefont {K.}~\bibnamefont
  {Hammerer}}, \bibinfo {author} {\bibfnamefont {M.~R.}\ \bibnamefont
  {Vanner}}, \ and\ \bibinfo {author} {\bibfnamefont {M.}~\bibnamefont
  {Aspelmeyer}},\ }\href {http://dx.doi.org/10.1038/nature08171} {\bibfield
  {journal} {\bibinfo  {journal} {Nature}\ }\textbf {\bibinfo {volume} {460}},\
  \bibinfo {pages} {724} (\bibinfo {year} {2009})}\BibitemShut {NoStop}%
\bibitem [{\citenamefont {Chan}\ \emph {et~al.}(2011)\citenamefont {Chan},
  \citenamefont {Alegre}, \citenamefont {Safavi-Naeini}, \citenamefont {Hill},
  \citenamefont {Krause}, \citenamefont {Groblacher}, \citenamefont
  {Aspelmeyer},\ and\ \citenamefont {Painter}}]{Chan_N478_89_11}%
  \BibitemOpen
  \bibfield  {author} {\bibinfo {author} {\bibfnamefont {J.}~\bibnamefont
  {Chan}}, \bibinfo {author} {\bibfnamefont {T.~P.~M.}\ \bibnamefont {Alegre}},
  \bibinfo {author} {\bibfnamefont {A.~H.}\ \bibnamefont {Safavi-Naeini}},
  \bibinfo {author} {\bibfnamefont {J.~T.}\ \bibnamefont {Hill}}, \bibinfo
  {author} {\bibfnamefont {A.}~\bibnamefont {Krause}}, \bibinfo {author}
  {\bibfnamefont {S.}~\bibnamefont {Groblacher}}, \bibinfo {author}
  {\bibfnamefont {M.}~\bibnamefont {Aspelmeyer}}, \ and\ \bibinfo {author}
  {\bibfnamefont {O.}~\bibnamefont {Painter}},\ }\href
  {http://dx.doi.org/10.1038/nature10461} {\bibfield  {journal} {\bibinfo
  {journal} {Nature}\ }\textbf {\bibinfo {volume} {478}},\ \bibinfo {pages}
  {89} (\bibinfo {year} {2011})}\BibitemShut {NoStop}%
\bibitem [{\citenamefont {Verhagen}\ \emph {et~al.}(2012)\citenamefont
  {Verhagen}, \citenamefont {Del{\'e}glise}, \citenamefont {Weis},
  \citenamefont {Schliesser},\ and\ \citenamefont
  {Kippenberg}}]{Verhagen_N482_63_12}%
  \BibitemOpen
  \bibfield  {author} {\bibinfo {author} {\bibfnamefont {E.}~\bibnamefont
  {Verhagen}}, \bibinfo {author} {\bibfnamefont {S.}~\bibnamefont
  {Del{\'e}glise}}, \bibinfo {author} {\bibfnamefont {S.}~\bibnamefont {Weis}},
  \bibinfo {author} {\bibfnamefont {A.}~\bibnamefont {Schliesser}}, \ and\
  \bibinfo {author} {\bibfnamefont {T.~J.}\ \bibnamefont {Kippenberg}},\ }\href
  {https://doi.org/10.1038/nature10787} {\bibfield  {journal} {\bibinfo
  {journal} {Nature}\ }\textbf {\bibinfo {volume} {482}},\ \bibinfo {pages}
  {63} (\bibinfo {year} {2012})}\BibitemShut {NoStop}%
\bibitem [{\citenamefont {Weis}\ \emph {et~al.}(2010)\citenamefont {Weis},
  \citenamefont {Rivi{\`e}re}, \citenamefont {Del{\'e}glise}, \citenamefont
  {Gavartin}, \citenamefont {Arcizet}, \citenamefont {Schliesser},\ and\
  \citenamefont {Kippenberg}}]{Weis_S330_1520_10}%
  \BibitemOpen
  \bibfield  {author} {\bibinfo {author} {\bibfnamefont {S.}~\bibnamefont
  {Weis}}, \bibinfo {author} {\bibfnamefont {R.}~\bibnamefont {Rivi{\`e}re}},
  \bibinfo {author} {\bibfnamefont {S.}~\bibnamefont {Del{\'e}glise}}, \bibinfo
  {author} {\bibfnamefont {E.}~\bibnamefont {Gavartin}}, \bibinfo {author}
  {\bibfnamefont {O.}~\bibnamefont {Arcizet}}, \bibinfo {author} {\bibfnamefont
  {A.}~\bibnamefont {Schliesser}}, \ and\ \bibinfo {author} {\bibfnamefont
  {T.~J.}\ \bibnamefont {Kippenberg}},\ }\href {\doibase
  10.1126/science.1195596} {\bibfield  {journal} {\bibinfo  {journal}
  {Science}\ }\textbf {\bibinfo {volume} {330}},\ \bibinfo {pages} {1520}
  (\bibinfo {year} {2010})},\ \Eprint
  {http://arxiv.org/abs/https://science.sciencemag.org/content/330/6010/1520.full.pdf}
  {https://science.sciencemag.org/content/330/6010/1520.full.pdf} \BibitemShut
  {NoStop}%
\bibitem [{\citenamefont {Weisbuch}\ \emph {et~al.}(1992)\citenamefont
  {Weisbuch}, \citenamefont {Nishioka}, \citenamefont {Ishikawa},\ and\
  \citenamefont {Arakawa}}]{Weisbuch92a}%
  \BibitemOpen
  \bibfield  {author} {\bibinfo {author} {\bibfnamefont {C.}~\bibnamefont
  {Weisbuch}}, \bibinfo {author} {\bibfnamefont {M.}~\bibnamefont {Nishioka}},
  \bibinfo {author} {\bibfnamefont {A.}~\bibnamefont {Ishikawa}}, \ and\
  \bibinfo {author} {\bibfnamefont {Y.}~\bibnamefont {Arakawa}},\ }\href
  {\doibase 10.1103/PhysRevLett.69.3314} {\bibfield  {journal} {\bibinfo
  {journal} {Phys. Rev. Lett.}\ }\textbf {\bibinfo {volume} {69}},\ \bibinfo
  {pages} {3314} (\bibinfo {year} {1992})}\BibitemShut {NoStop}%
\bibitem [{\citenamefont {Kasprzak}\ \emph {et~al.}(2006)\citenamefont
  {Kasprzak}, \citenamefont {Richard}, \citenamefont {Kundermann},
  \citenamefont {Baas}, \citenamefont {Jeambrun}, \citenamefont {Keeling},
  \citenamefont {Marchetti}, \citenamefont {Szyma\'nska}, \citenamefont
  {Andr{\'e}}, \citenamefont {Staehli}, \citenamefont {Savona}, \citenamefont
  {Littlewood}, \citenamefont {Deveaud},\ and\ \citenamefont
  {Dang}}]{Kasprzak_N443_409_06}%
  \BibitemOpen
  \bibfield  {author} {\bibinfo {author} {\bibfnamefont {J.}~\bibnamefont
  {Kasprzak}}, \bibinfo {author} {\bibfnamefont {M.}~\bibnamefont {Richard}},
  \bibinfo {author} {\bibfnamefont {S.}~\bibnamefont {Kundermann}}, \bibinfo
  {author} {\bibfnamefont {A.}~\bibnamefont {Baas}}, \bibinfo {author}
  {\bibfnamefont {P.}~\bibnamefont {Jeambrun}}, \bibinfo {author}
  {\bibfnamefont {J.~M.~J.}\ \bibnamefont {Keeling}}, \bibinfo {author}
  {\bibfnamefont {F.~M.}\ \bibnamefont {Marchetti}}, \bibinfo {author}
  {\bibfnamefont {M.~H.}\ \bibnamefont {Szyma\'nska}}, \bibinfo {author}
  {\bibfnamefont {R.}~\bibnamefont {Andr{\'e}}}, \bibinfo {author}
  {\bibfnamefont {J.~L.}\ \bibnamefont {Staehli}}, \bibinfo {author}
  {\bibfnamefont {V.}~\bibnamefont {Savona}}, \bibinfo {author} {\bibfnamefont
  {P.~B.}\ \bibnamefont {Littlewood}}, \bibinfo {author} {\bibfnamefont
  {B.}~\bibnamefont {Deveaud}}, \ and\ \bibinfo {author} {\bibfnamefont
  {L.~S.}\ \bibnamefont {Dang}},\ }\href {https://doi.org/10.1038/nature05131}
  {\bibfield  {journal} {\bibinfo  {journal} {Nature}\ }\textbf {\bibinfo
  {volume} {443}},\ \bibinfo {pages} {409} (\bibinfo {year}
  {2006})}\BibitemShut {NoStop}%
\bibitem [{\citenamefont {Sanvitto}\ and\ \citenamefont
  {Kena-Cohen}(2016)}]{Sanvitto_NM15_1061_16}%
  \BibitemOpen
  \bibfield  {author} {\bibinfo {author} {\bibfnamefont {D.}~\bibnamefont
  {Sanvitto}}\ and\ \bibinfo {author} {\bibfnamefont {S.}~\bibnamefont
  {Kena-Cohen}},\ }\href {http://dx.doi.org/10.1038/nmat4668} {\bibfield
  {journal} {\bibinfo  {journal} {Nat. Mater.}\ }\textbf {\bibinfo {volume}
  {15}},\ \bibinfo {pages} {1061} (\bibinfo {year} {2016})}\BibitemShut
  {NoStop}%
\bibitem [{\citenamefont {{de Lima, Jr.}}\ and\ \citenamefont
  {Santos}(2005)}]{PVS156}%
  \BibitemOpen
  \bibfield  {author} {\bibinfo {author} {\bibfnamefont {M.~M.}\ \bibnamefont
  {{de Lima, Jr.}}}\ and\ \bibinfo {author} {\bibfnamefont {P.~V.}\
  \bibnamefont {Santos}},\ }\href {\doibase 10.1088/0034-4885/68/7/r02}
  {\bibfield  {journal} {\bibinfo  {journal} {Rep. Prog. Phys.}\ }\textbf
  {\bibinfo {volume} {68}},\ \bibinfo {pages} {1639} (\bibinfo {year}
  {2005})}\BibitemShut {NoStop}%
\bibitem [{\citenamefont {de~Lima}\ \emph {et~al.}(2006)\citenamefont
  {de~Lima}, \citenamefont {van~der Poel}, \citenamefont {Santos},\ and\
  \citenamefont {Hvam}}]{PVS169}%
  \BibitemOpen
  \bibfield  {author} {\bibinfo {author} {\bibfnamefont {M.~M.}\ \bibnamefont
  {de~Lima}}, \bibinfo {author} {\bibfnamefont {M.}~\bibnamefont {van~der
  Poel}}, \bibinfo {author} {\bibfnamefont {P.~V.}\ \bibnamefont {Santos}}, \
  and\ \bibinfo {author} {\bibfnamefont {J.~M.}\ \bibnamefont {Hvam}},\ }\href
  {\doibase 10.1103/PhysRevLett.97.045501} {\bibfield  {journal} {\bibinfo
  {journal} {Phys. Rev. Lett.}\ }\textbf {\bibinfo {volume} {97}},\ \bibinfo
  {pages} {045501} (\bibinfo {year} {2006})}\BibitemShut {NoStop}%
\bibitem [{\citenamefont {Berstermann}\ \emph {et~al.}(2012)\citenamefont
  {Berstermann}, \citenamefont {Br{\"u}ggemann}, \citenamefont {Akimov},
  \citenamefont {Bombeck}, \citenamefont {Yakovlev}, \citenamefont {Gippius},
  \citenamefont {Scherbakov}, \citenamefont {Sagnes}, \citenamefont {Bloch},\
  and\ \citenamefont {Bayer}}]{Berstermann_PRB86_195306_12}%
  \BibitemOpen
  \bibfield  {author} {\bibinfo {author} {\bibfnamefont {T.}~\bibnamefont
  {Berstermann}}, \bibinfo {author} {\bibfnamefont {C.}~\bibnamefont
  {Br{\"u}ggemann}}, \bibinfo {author} {\bibfnamefont {A.~V.}\ \bibnamefont
  {Akimov}}, \bibinfo {author} {\bibfnamefont {M.}~\bibnamefont {Bombeck}},
  \bibinfo {author} {\bibfnamefont {D.~R.}\ \bibnamefont {Yakovlev}}, \bibinfo
  {author} {\bibfnamefont {N.~A.}\ \bibnamefont {Gippius}}, \bibinfo {author}
  {\bibfnamefont {A.~V.}\ \bibnamefont {Scherbakov}}, \bibinfo {author}
  {\bibfnamefont {I.}~\bibnamefont {Sagnes}}, \bibinfo {author} {\bibfnamefont
  {J.}~\bibnamefont {Bloch}}, \ and\ \bibinfo {author} {\bibfnamefont
  {M.}~\bibnamefont {Bayer}},\ }\href {\doibase 10.1103/PhysRevB.86.195306}
  {\bibfield  {journal} {\bibinfo  {journal} {Phys. Rev. B}\ }\textbf {\bibinfo
  {volume} {86}},\ \bibinfo {pages} {195306} (\bibinfo {year}
  {2012})}\BibitemShut {NoStop}%
\bibitem [{\citenamefont {Berstermann}\ \emph {et~al.}(2009)\citenamefont
  {Berstermann}, \citenamefont {Scherbakov}, \citenamefont {Akimov},
  \citenamefont {Yakovlev}, \citenamefont {Gippius}, \citenamefont {Glavin},
  \citenamefont {Sagnes}, \citenamefont {Bloch},\ and\ \citenamefont
  {Bayer}}]{Berstermann_PRB80_75301_09}%
  \BibitemOpen
  \bibfield  {author} {\bibinfo {author} {\bibfnamefont {T.}~\bibnamefont
  {Berstermann}}, \bibinfo {author} {\bibfnamefont {A.~V.}\ \bibnamefont
  {Scherbakov}}, \bibinfo {author} {\bibfnamefont {A.~V.}\ \bibnamefont
  {Akimov}}, \bibinfo {author} {\bibfnamefont {D.~R.}\ \bibnamefont
  {Yakovlev}}, \bibinfo {author} {\bibfnamefont {N.~A.}\ \bibnamefont
  {Gippius}}, \bibinfo {author} {\bibfnamefont {B.~A.}\ \bibnamefont {Glavin}},
  \bibinfo {author} {\bibfnamefont {I.}~\bibnamefont {Sagnes}}, \bibinfo
  {author} {\bibfnamefont {J.}~\bibnamefont {Bloch}}, \ and\ \bibinfo {author}
  {\bibfnamefont {M.}~\bibnamefont {Bayer}},\ }\href {\doibase
  10.1103/PhysRevB.80.075301} {\bibfield  {journal} {\bibinfo  {journal} {Phys.
  Rev. B}\ }\textbf {\bibinfo {volume} {80}},\ \bibinfo {pages} {075301}
  (\bibinfo {year} {2009})}\BibitemShut {NoStop}%
\bibitem [{\citenamefont {Metcalfe}\ \emph {et~al.}(2010)\citenamefont
  {Metcalfe}, \citenamefont {Carr}, \citenamefont {Muller}, \citenamefont
  {Solomon},\ and\ \citenamefont {Lawall}}]{Metcalfe_PRL105_37401_10}%
  \BibitemOpen
  \bibfield  {author} {\bibinfo {author} {\bibfnamefont {M.}~\bibnamefont
  {Metcalfe}}, \bibinfo {author} {\bibfnamefont {S.~M.}\ \bibnamefont {Carr}},
  \bibinfo {author} {\bibfnamefont {A.}~\bibnamefont {Muller}}, \bibinfo
  {author} {\bibfnamefont {G.~S.}\ \bibnamefont {Solomon}}, \ and\ \bibinfo
  {author} {\bibfnamefont {J.}~\bibnamefont {Lawall}},\ }\href {\doibase
  10.1103/PhysRevLett.105.037401} {\bibfield  {journal} {\bibinfo  {journal}
  {Phys. Rev. Lett.}\ }\textbf {\bibinfo {volume} {105}},\ \bibinfo {pages}
  {037401} (\bibinfo {year} {2010})}\BibitemShut {NoStop}%
\bibitem [{\citenamefont {Trigo}\ \emph {et~al.}(2002)\citenamefont {Trigo},
  \citenamefont {Bruchhausen}, \citenamefont {Fainstein}, \citenamefont
  {Jusserand},\ and\ \citenamefont {Thierry-Mieg}}]{Trigo02a}%
  \BibitemOpen
  \bibfield  {author} {\bibinfo {author} {\bibfnamefont {M.}~\bibnamefont
  {Trigo}}, \bibinfo {author} {\bibfnamefont {A.}~\bibnamefont {Bruchhausen}},
  \bibinfo {author} {\bibfnamefont {A.}~\bibnamefont {Fainstein}}, \bibinfo
  {author} {\bibfnamefont {B.}~\bibnamefont {Jusserand}}, \ and\ \bibinfo
  {author} {\bibfnamefont {V.}~\bibnamefont {Thierry-Mieg}},\ }\href {\doibase
  10.1103/PhysRevLett.89.227402} {\bibfield  {journal} {\bibinfo  {journal}
  {Phys. Rev. Lett.}\ }\textbf {\bibinfo {volume} {89}},\ \bibinfo {pages}
  {227402} (\bibinfo {year} {2002})}\BibitemShut {NoStop}%
\bibitem [{\citenamefont {Fainstein}\ \emph {et~al.}(2013)\citenamefont
  {Fainstein}, \citenamefont {Lanzillotti-Kimura}, \citenamefont {Jusserand},\
  and\ \citenamefont {Perrin}}]{Fainstein_PRL110_37403_13}%
  \BibitemOpen
  \bibfield  {author} {\bibinfo {author} {\bibfnamefont {A.}~\bibnamefont
  {Fainstein}}, \bibinfo {author} {\bibfnamefont {N.~D.}\ \bibnamefont
  {Lanzillotti-Kimura}}, \bibinfo {author} {\bibfnamefont {B.}~\bibnamefont
  {Jusserand}}, \ and\ \bibinfo {author} {\bibfnamefont {B.}~\bibnamefont
  {Perrin}},\ }\href {\doibase 10.1103/PhysRevLett.110.037403} {\bibfield
  {journal} {\bibinfo  {journal} {Phys. Rev. Lett.}\ }\textbf {\bibinfo
  {volume} {110}},\ \bibinfo {pages} {037403} (\bibinfo {year}
  {2013})}\BibitemShut {NoStop}%
\bibitem [{\citenamefont {Kyriienko}\ \emph {et~al.}(2014)\citenamefont
  {Kyriienko}, \citenamefont {Liew},\ and\ \citenamefont
  {Shelykh}}]{Kyriienko_PRL112_76402_14}%
  \BibitemOpen
  \bibfield  {author} {\bibinfo {author} {\bibfnamefont {O.}~\bibnamefont
  {Kyriienko}}, \bibinfo {author} {\bibfnamefont {T.~C.~H.}\ \bibnamefont
  {Liew}}, \ and\ \bibinfo {author} {\bibfnamefont {I.~A.}\ \bibnamefont
  {Shelykh}},\ }\href {\doibase 10.1103/PhysRevLett.112.076402} {\bibfield
  {journal} {\bibinfo  {journal} {Phys. Rev. Lett.}\ }\textbf {\bibinfo
  {volume} {112}},\ \bibinfo {pages} {076402} (\bibinfo {year}
  {2014})}\BibitemShut {NoStop}%
\bibitem [{\citenamefont {Rozas}\ \emph {et~al.}(2014)\citenamefont {Rozas},
  \citenamefont {Bruchhausen}, \citenamefont {Fainstein}, \citenamefont
  {Jusserand},\ and\ \citenamefont {Lema{\^i}tre}}]{Rozas_PRB90_201302_14}%
  \BibitemOpen
  \bibfield  {author} {\bibinfo {author} {\bibfnamefont {G.}~\bibnamefont
  {Rozas}}, \bibinfo {author} {\bibfnamefont {A.~E.}\ \bibnamefont
  {Bruchhausen}}, \bibinfo {author} {\bibfnamefont {A.}~\bibnamefont
  {Fainstein}}, \bibinfo {author} {\bibfnamefont {B.}~\bibnamefont
  {Jusserand}}, \ and\ \bibinfo {author} {\bibfnamefont {A.}~\bibnamefont
  {Lema{\^i}tre}},\ }\href {\doibase 10.1103/PhysRevB.90.201302} {\bibfield
  {journal} {\bibinfo  {journal} {Phys. Rev. B}\ }\textbf {\bibinfo {volume}
  {90}},\ \bibinfo {pages} {201302} (\bibinfo {year} {2014})}\BibitemShut
  {NoStop}%
\bibitem [{\citenamefont {Jusserand}\ \emph {et~al.}(2015)\citenamefont
  {Jusserand}, \citenamefont {Poddubny}, \citenamefont {Poshakinskiy},
  \citenamefont {Fainstein},\ and\ \citenamefont
  {Lemaitre}}]{Jusserand_PRL115_267402_15}%
  \BibitemOpen
  \bibfield  {author} {\bibinfo {author} {\bibfnamefont {B.}~\bibnamefont
  {Jusserand}}, \bibinfo {author} {\bibfnamefont {A.~N.}\ \bibnamefont
  {Poddubny}}, \bibinfo {author} {\bibfnamefont {A.~V.}\ \bibnamefont
  {Poshakinskiy}}, \bibinfo {author} {\bibfnamefont {A.}~\bibnamefont
  {Fainstein}}, \ and\ \bibinfo {author} {\bibfnamefont {A.}~\bibnamefont
  {Lemaitre}},\ }\href {\doibase 10.1103/PhysRevLett.115.267402} {\bibfield
  {journal} {\bibinfo  {journal} {Phys. Rev. Lett.}\ }\textbf {\bibinfo
  {volume} {115}},\ \bibinfo {pages} {267402} (\bibinfo {year}
  {2015})}\BibitemShut {NoStop}%
\bibitem [{\citenamefont {Machado}\ \emph {et~al.}(2019)\citenamefont
  {Machado}, \citenamefont {Crespo-Poveda}, \citenamefont {Kuznetsov},
  \citenamefont {Biermann}, \citenamefont {Scalvi},\ and\ \citenamefont
  {Santos}}]{PVS327}%
  \BibitemOpen
  \bibfield  {author} {\bibinfo {author} {\bibfnamefont {D.~H.}\ \bibnamefont
  {Machado}}, \bibinfo {author} {\bibfnamefont {A.}~\bibnamefont
  {Crespo-Poveda}}, \bibinfo {author} {\bibfnamefont {A.~S.}\ \bibnamefont
  {Kuznetsov}}, \bibinfo {author} {\bibfnamefont {K.}~\bibnamefont {Biermann}},
  \bibinfo {author} {\bibfnamefont {L.~V.}\ \bibnamefont {Scalvi}}, \ and\
  \bibinfo {author} {\bibfnamefont {P.~V.}\ \bibnamefont {Santos}},\ }\href
  {\doibase 10.1103/PhysRevApplied.12.044013} {\bibfield  {journal} {\bibinfo
  {journal} {Phys. Rev. Applied}\ }\textbf {\bibinfo {volume} {12}},\ \bibinfo
  {pages} {044013} (\bibinfo {year} {2019})},\ \bibinfo {note}
  {arXiv:1907.09787v2}\BibitemShut {NoStop}%
\bibitem [{\citenamefont {Cerda-M{\'e}ndez}\ \emph {et~al.}(2010)\citenamefont
  {Cerda-M{\'e}ndez}, \citenamefont {Krizhanovskii}, \citenamefont {Wouters},
  \citenamefont {Bradley}, \citenamefont {Biermann}, \citenamefont {Guda},
  \citenamefont {Hey}, \citenamefont {Santos}, \citenamefont {Sarkar},\ and\
  \citenamefont {Skolnick}}]{PVS223}%
  \BibitemOpen
  \bibfield  {author} {\bibinfo {author} {\bibfnamefont {E.~A.}\ \bibnamefont
  {Cerda-M{\'e}ndez}}, \bibinfo {author} {\bibfnamefont {D.~N.}\ \bibnamefont
  {Krizhanovskii}}, \bibinfo {author} {\bibfnamefont {M.}~\bibnamefont
  {Wouters}}, \bibinfo {author} {\bibfnamefont {R.}~\bibnamefont {Bradley}},
  \bibinfo {author} {\bibfnamefont {K.}~\bibnamefont {Biermann}}, \bibinfo
  {author} {\bibfnamefont {K.}~\bibnamefont {Guda}}, \bibinfo {author}
  {\bibfnamefont {R.}~\bibnamefont {Hey}}, \bibinfo {author} {\bibfnamefont
  {P.~V.}\ \bibnamefont {Santos}}, \bibinfo {author} {\bibfnamefont
  {D.}~\bibnamefont {Sarkar}}, \ and\ \bibinfo {author} {\bibfnamefont {M.~S.}\
  \bibnamefont {Skolnick}},\ }\href {\doibase 10.1103/PhysRevLett.105.116402}
  {\bibfield  {journal} {\bibinfo  {journal} {Phys. Rev. Lett.}\ }\textbf
  {\bibinfo {volume} {105}},\ \bibinfo {pages} {116402} (\bibinfo {year}
  {2010})}\BibitemShut {NoStop}%
\bibitem [{\citenamefont {Dular}\ \emph {et~al.}(1998)\citenamefont {Dular},
  \citenamefont {Geuzaine}, \citenamefont {Henrotte},\ and\ \citenamefont
  {Legros}}]{getdp-ieee1998}%
  \BibitemOpen
  \bibfield  {author} {\bibinfo {author} {\bibfnamefont {P.}~\bibnamefont
  {Dular}}, \bibinfo {author} {\bibfnamefont {C.}~\bibnamefont {Geuzaine}},
  \bibinfo {author} {\bibfnamefont {F.}~\bibnamefont {Henrotte}}, \ and\
  \bibinfo {author} {\bibfnamefont {W.}~\bibnamefont {Legros}},\ }\href
  {https://doi.org/10.1109/20.717799} {\bibfield  {journal} {\bibinfo
  {journal} {{IEEE} Trans. Magn.}\ }\textbf {\bibinfo {volume} {34}},\ \bibinfo
  {pages} {3395} (\bibinfo {year} {1998})}\BibitemShut {NoStop}%
\bibitem [{\citenamefont {Sogawa}\ \emph {et~al.}(2001)\citenamefont {Sogawa},
  \citenamefont {Santos}, \citenamefont {Zhang}, \citenamefont {Eshlaghi},
  \citenamefont {Wieck},\ and\ \citenamefont {Ploog}}]{PVS107}%
  \BibitemOpen
  \bibfield  {author} {\bibinfo {author} {\bibfnamefont {T.}~\bibnamefont
  {Sogawa}}, \bibinfo {author} {\bibfnamefont {P.~V.}\ \bibnamefont {Santos}},
  \bibinfo {author} {\bibfnamefont {S.~K.}\ \bibnamefont {Zhang}}, \bibinfo
  {author} {\bibfnamefont {S.}~\bibnamefont {Eshlaghi}}, \bibinfo {author}
  {\bibfnamefont {A.~D.}\ \bibnamefont {Wieck}}, \ and\ \bibinfo {author}
  {\bibfnamefont {K.~H.}\ \bibnamefont {Ploog}},\ }\href {\doibase
  10.1103/PhysRevB.63.121307} {\bibfield  {journal} {\bibinfo  {journal} {Phys.
  Rev. B}\ }\textbf {\bibinfo {volume} {63}},\ \bibinfo {pages} {121307}
  (\bibinfo {year} {2001})}\BibitemShut {NoStop}%
\bibitem [{\citenamefont {Madelung}(1982)}]{LB17a}%
  \BibitemOpen
  \bibinfo {editor} {\bibfnamefont {O.}~\bibnamefont {Madelung}},\ ed.,\
  \href@noop {} {\emph {\bibinfo {title} {Landolt-B\"{o}rnstein}}},\ Vol.\
  \bibinfo {volume} {17a}\ (\bibinfo  {publisher} {Springer Verlag},\ \bibinfo
  {address} {London},\ \bibinfo {year} {1982})\BibitemShut {NoStop}%
\bibitem [{\citenamefont {Iikawa}\ \emph {et~al.}(2019)\citenamefont {Iikawa},
  \citenamefont {Hern{\'a}ndez-M{\'i}nguez}, \citenamefont {Aharonovich},
  \citenamefont {Nakhaie}, \citenamefont {Liou}, \citenamefont {Lopes},\ and\
  \citenamefont {Santos}}]{PVS323}%
  \BibitemOpen
  \bibfield  {author} {\bibinfo {author} {\bibfnamefont {F.}~\bibnamefont
  {Iikawa}}, \bibinfo {author} {\bibfnamefont {A.}~\bibnamefont
  {Hern{\'a}ndez-M{\'i}nguez}}, \bibinfo {author} {\bibfnamefont
  {I.}~\bibnamefont {Aharonovich}}, \bibinfo {author} {\bibfnamefont
  {S.}~\bibnamefont {Nakhaie}}, \bibinfo {author} {\bibfnamefont {Y.-T.}\
  \bibnamefont {Liou}}, \bibinfo {author} {\bibfnamefont {J.~M.~J.}\
  \bibnamefont {Lopes}}, \ and\ \bibinfo {author} {\bibfnamefont {P.~V.}\
  \bibnamefont {Santos}},\ }\href {\doibase 10.1063/1.5093299} {\bibfield
  {journal} {\bibinfo  {journal} {Appl. Phys. Lett.}\ }\textbf {\bibinfo
  {volume} {171104}},\ \bibinfo {pages} {171104} (\bibinfo {year}
  {2019})}\BibitemShut {NoStop}%
\bibitem [{\citenamefont {Cerda-M\'endez}\ \emph {et~al.}(2012)\citenamefont
  {Cerda-M\'endez}, \citenamefont {Krizhanovskii}, \citenamefont {Biermann},
  \citenamefont {Hey}, \citenamefont {Skolnick},\ and\ \citenamefont
  {Santos}}]{PVS255}%
  \BibitemOpen
  \bibfield  {author} {\bibinfo {author} {\bibfnamefont {E.~A.}\ \bibnamefont
  {Cerda-M\'endez}}, \bibinfo {author} {\bibfnamefont {D.~N.}\ \bibnamefont
  {Krizhanovskii}}, \bibinfo {author} {\bibfnamefont {K.}~\bibnamefont
  {Biermann}}, \bibinfo {author} {\bibfnamefont {R.}~\bibnamefont {Hey}},
  \bibinfo {author} {\bibfnamefont {M.~S.}\ \bibnamefont {Skolnick}}, \ and\
  \bibinfo {author} {\bibfnamefont {P.~V.}\ \bibnamefont {Santos}},\ }\href
  {http://stacks.iop.org/1367-2630/14/i=7/a=075011} {\bibfield  {journal}
  {\bibinfo  {journal} {New J. Phys.}\ }\textbf {\bibinfo {volume} {14}},\
  \bibinfo {pages} {075011} (\bibinfo {year} {2012})}\BibitemShut {NoStop}%
\bibitem [{\citenamefont {Chafatinos}\ \emph {et~al.}(2020)\citenamefont
  {Chafatinos}, \citenamefont {Kuznetsov}, \citenamefont {Anguiano},
  \citenamefont {Bruchhausen}, \citenamefont {Reynoso}, \citenamefont
  {Biermann}, \citenamefont {Santos},\ and\ \citenamefont
  {Fainstein}}]{PVS333}%
  \BibitemOpen
  \bibfield  {author} {\bibinfo {author} {\bibfnamefont {D.~L.}\ \bibnamefont
  {Chafatinos}}, \bibinfo {author} {\bibfnamefont {A.~S.}\ \bibnamefont
  {Kuznetsov}}, \bibinfo {author} {\bibfnamefont {S.}~\bibnamefont {Anguiano}},
  \bibinfo {author} {\bibfnamefont {A.~E.}\ \bibnamefont {Bruchhausen}},
  \bibinfo {author} {\bibfnamefont {A.~A.}\ \bibnamefont {Reynoso}}, \bibinfo
  {author} {\bibfnamefont {K.}~\bibnamefont {Biermann}}, \bibinfo {author}
  {\bibfnamefont {P.~V.}\ \bibnamefont {Santos}}, \ and\ \bibinfo {author}
  {\bibfnamefont {A.}~\bibnamefont {Fainstein}},\ }\href@noop {} {\bibfield
  {journal} {\bibinfo  {journal} {arXiv:2001.09958v1}\ } (\bibinfo {year}
  {2020})}\BibitemShut {NoStop}%
\bibitem [{\citenamefont {Hamoumi}\ \emph {et~al.}(2018)\citenamefont
  {Hamoumi}, \citenamefont {Allain}, \citenamefont {Hease}, \citenamefont
  {Gil-Santos}, \citenamefont {Morgenroth}, \citenamefont {G\'erard},
  \citenamefont {Lema\^{\i}tre}, \citenamefont {Leo},\ and\ \citenamefont
  {Favero}}]{Hamoumi_PRL120_223601_18}%
  \BibitemOpen
  \bibfield  {author} {\bibinfo {author} {\bibfnamefont {M.}~\bibnamefont
  {Hamoumi}}, \bibinfo {author} {\bibfnamefont {P.~E.}\ \bibnamefont {Allain}},
  \bibinfo {author} {\bibfnamefont {W.}~\bibnamefont {Hease}}, \bibinfo
  {author} {\bibfnamefont {E.}~\bibnamefont {Gil-Santos}}, \bibinfo {author}
  {\bibfnamefont {L.}~\bibnamefont {Morgenroth}}, \bibinfo {author}
  {\bibfnamefont {B.}~\bibnamefont {G\'erard}}, \bibinfo {author}
  {\bibfnamefont {A.}~\bibnamefont {Lema\^{\i}tre}}, \bibinfo {author}
  {\bibfnamefont {G.}~\bibnamefont {Leo}}, \ and\ \bibinfo {author}
  {\bibfnamefont {I.}~\bibnamefont {Favero}},\ }\href {\doibase
  10.1103/PhysRevLett.120.223601} {\bibfield  {journal} {\bibinfo  {journal}
  {Phys. Rev. Lett.}\ }\textbf {\bibinfo {volume} {120}},\ \bibinfo {pages}
  {223601} (\bibinfo {year} {2018})}\BibitemShut {NoStop}%
\bibitem [{\citenamefont {Schneider}\ \emph {et~al.}(2017)\citenamefont
  {Schneider}, \citenamefont {Winkler}, \citenamefont {Fraser}, \citenamefont
  {Kamp}, \citenamefont {Yamamoto}, \citenamefont {Ostrovskaya},\ and\
  \citenamefont {H{\"o}fling}}]{Schneider_RPP_16503_17}%
  \BibitemOpen
  \bibfield  {author} {\bibinfo {author} {\bibfnamefont {C.}~\bibnamefont
  {Schneider}}, \bibinfo {author} {\bibfnamefont {K.}~\bibnamefont {Winkler}},
  \bibinfo {author} {\bibfnamefont {M.~D.}\ \bibnamefont {Fraser}}, \bibinfo
  {author} {\bibfnamefont {M.}~\bibnamefont {Kamp}}, \bibinfo {author}
  {\bibfnamefont {Y.}~\bibnamefont {Yamamoto}}, \bibinfo {author}
  {\bibfnamefont {E.~A.}\ \bibnamefont {Ostrovskaya}}, \ and\ \bibinfo {author}
  {\bibfnamefont {S.}~\bibnamefont {H{\"o}fling}},\ }\href {\doibase
  10.1088/0034-4885/80/1/016503} {\bibfield  {journal} {\bibinfo  {journal}
  {Rep. Prog. Phys.}\ }\textbf {\bibinfo {volume} {80}},\ \bibinfo {pages}
  {016503} (\bibinfo {year} {2017})}\BibitemShut {NoStop}%
\bibitem [{\citenamefont {Kuznetsov}\ \emph {et~al.}(2018)\citenamefont
  {Kuznetsov}, \citenamefont {Helgers}, \citenamefont {Biermann},\ and\
  \citenamefont {Santos}}]{PVS312}%
  \BibitemOpen
  \bibfield  {author} {\bibinfo {author} {\bibfnamefont {A.~S.}\ \bibnamefont
  {Kuznetsov}}, \bibinfo {author} {\bibfnamefont {P.~L.~J.}\ \bibnamefont
  {Helgers}}, \bibinfo {author} {\bibfnamefont {K.}~\bibnamefont {Biermann}}, \
  and\ \bibinfo {author} {\bibfnamefont {P.~V.}\ \bibnamefont {Santos}},\
  }\href {\doibase 10.1103/PhysRevB.97.195309} {\bibfield  {journal} {\bibinfo
  {journal} {Phys. Rev. B}\ }\textbf {\bibinfo {volume} {97}},\ \bibinfo
  {pages} {195309} (\bibinfo {year} {2018})}\BibitemShut {NoStop}%
\bibitem [{\citenamefont {Anguiano}\ \emph {et~al.}(2017)\citenamefont
  {Anguiano}, \citenamefont {Bruchhausen}, \citenamefont {Jusserand},
  \citenamefont {Favero}, \citenamefont {Lamberti}, \citenamefont {Lanco},
  \citenamefont {Sagnes}, \citenamefont {Lema\^{\i}tre}, \citenamefont
  {Lanzillotti-Kimura}, \citenamefont {Senellart},\ and\ \citenamefont
  {Fainstein}}]{Anguiano_PRL118_263901_17}%
  \BibitemOpen
  \bibfield  {author} {\bibinfo {author} {\bibfnamefont {S.}~\bibnamefont
  {Anguiano}}, \bibinfo {author} {\bibfnamefont {A.~E.}\ \bibnamefont
  {Bruchhausen}}, \bibinfo {author} {\bibfnamefont {B.}~\bibnamefont
  {Jusserand}}, \bibinfo {author} {\bibfnamefont {I.}~\bibnamefont {Favero}},
  \bibinfo {author} {\bibfnamefont {F.~R.}\ \bibnamefont {Lamberti}}, \bibinfo
  {author} {\bibfnamefont {L.}~\bibnamefont {Lanco}}, \bibinfo {author}
  {\bibfnamefont {I.}~\bibnamefont {Sagnes}}, \bibinfo {author} {\bibfnamefont
  {A.}~\bibnamefont {Lema\^{\i}tre}}, \bibinfo {author} {\bibfnamefont {N.~D.}\
  \bibnamefont {Lanzillotti-Kimura}}, \bibinfo {author} {\bibfnamefont
  {P.}~\bibnamefont {Senellart}}, \ and\ \bibinfo {author} {\bibfnamefont
  {A.}~\bibnamefont {Fainstein}},\ }\href {\doibase
  10.1103/PhysRevLett.118.263901} {\bibfield  {journal} {\bibinfo  {journal}
  {Phys. Rev. Lett.}\ }\textbf {\bibinfo {volume} {118}},\ \bibinfo {pages}
  {263901} (\bibinfo {year} {2017})}\BibitemShut {NoStop}%
\bibitem [{\citenamefont {{Lakin}}(2005)}]{Lakin_ITUFFC52_707_05}%
  \BibitemOpen
  \bibfield  {author} {\bibinfo {author} {\bibfnamefont {K.~M.}\ \bibnamefont
  {{Lakin}}},\ }\href {\doibase 10.1109/TUFFC.2005.1503959} {\bibfield
  {journal} {\bibinfo  {journal} {IEEE Transactions on Ultrasonics,
  Ferroelectrics, and Frequency Control}\ }\textbf {\bibinfo {volume} {52}},\
  \bibinfo {pages} {707} (\bibinfo {year} {2005})}\BibitemShut {NoStop}%
\bibitem [{\citenamefont {Iikawa}\ \emph {et~al.}(2016)\citenamefont {Iikawa},
  \citenamefont {Hern{\'a}ndez-M{\'i}nguez}, \citenamefont {Ramsteiner},\ and\
  \citenamefont {Santos}}]{PVS282}%
  \BibitemOpen
  \bibfield  {author} {\bibinfo {author} {\bibfnamefont {F.}~\bibnamefont
  {Iikawa}}, \bibinfo {author} {\bibfnamefont {A.}~\bibnamefont
  {Hern{\'a}ndez-M{\'i}nguez}}, \bibinfo {author} {\bibfnamefont
  {M.}~\bibnamefont {Ramsteiner}}, \ and\ \bibinfo {author} {\bibfnamefont
  {P.~V.}\ \bibnamefont {Santos}},\ }\href {\doibase
  10.1103/PhysRevB.93.195212} {\bibfield  {journal} {\bibinfo  {journal} {Phys.
  Rev. B}\ }\textbf {\bibinfo {volume} {93}},\ \bibinfo {pages} {195212}
  (\bibinfo {year} {2016})}\BibitemShut {NoStop}%
\end{thebibliography}%





\clearpage

\widetext 

\section*{Supplementary Material}

\vspace{1cm }
\centerline{\large \bf Electrically driven exciton-polariton optomechanics at super high frequencies}

\vspace{.25cm }

\centerline{ \bf Alexander S. Kuznetsov,$^1$ Diego H. O. Machado,$^{1,2}$ Klaus Biermann,$^1$ and Paulo V. Santos$^1$}

\vspace{.25cm }

\noindent $^1$Paul-Drude-Institut f{\"u}r Festk{\"o}rperelektronik, Leibniz-Institut im Forschungsverbund Berlin e. V., Hausvogteiplatz 5-7, 10117 Berlin, Germany \\
\noindent $^{2}$UNESP, S{\~a}o Paulo State University, Department of Physics,
Av. Eng. Luiz Edmundo C. Coube 14-01, 17033-360, Bauru, SP, Brazil.

\vspace{1cm }

\beginsupplement

\section{Field distribution in hybrid microcavities}
\label{Field distribution in hybrid microcavities}

The modulation of the quantum well (QW) energy levels by the strain field $u_{zz}=\partial u/\partial z$ of the BAW is the dominating mechanism determining the opto-mechanical coupling in these structures. Here, ${\bf u}=(0,0,u_z)$ is the BAW displacement field as a function of the $z$ coordinate perpendicular to the MC surface.  Optimization of this coupling requires, therefore, that the QWs embedded in the microcavity (MC) spacer are placed close to the anti-nodes of both the optical and the acoustic strain fields.

It turns out that the requirement stated above can never be strictly satisfied.  
In Al\textsubscript{x}Ga\textsubscript{1-x}As alloys,  the  coincident ratio between the light and sound velocities as well as between the inverse acoustic impedance is approximately independent of the composition $x$ mentioned in the main text also implies that the anti-nodes of the acoustic ($u_z$ ) and optical field ($F_x$, assumed to be polarized along the surface direction $x$) occur at the same $z$ coordinate~\cite{Fainstein_PRL110_37403_13}. The anti-nodes of $u_z$ are, however, nodes of $u_{zz}$, thus implying in a vanishing modulation of the excitonic energy.

\begin{figure}[!bthp]
\includegraphics[width=0.8\textwidth, keepaspectratio=true]{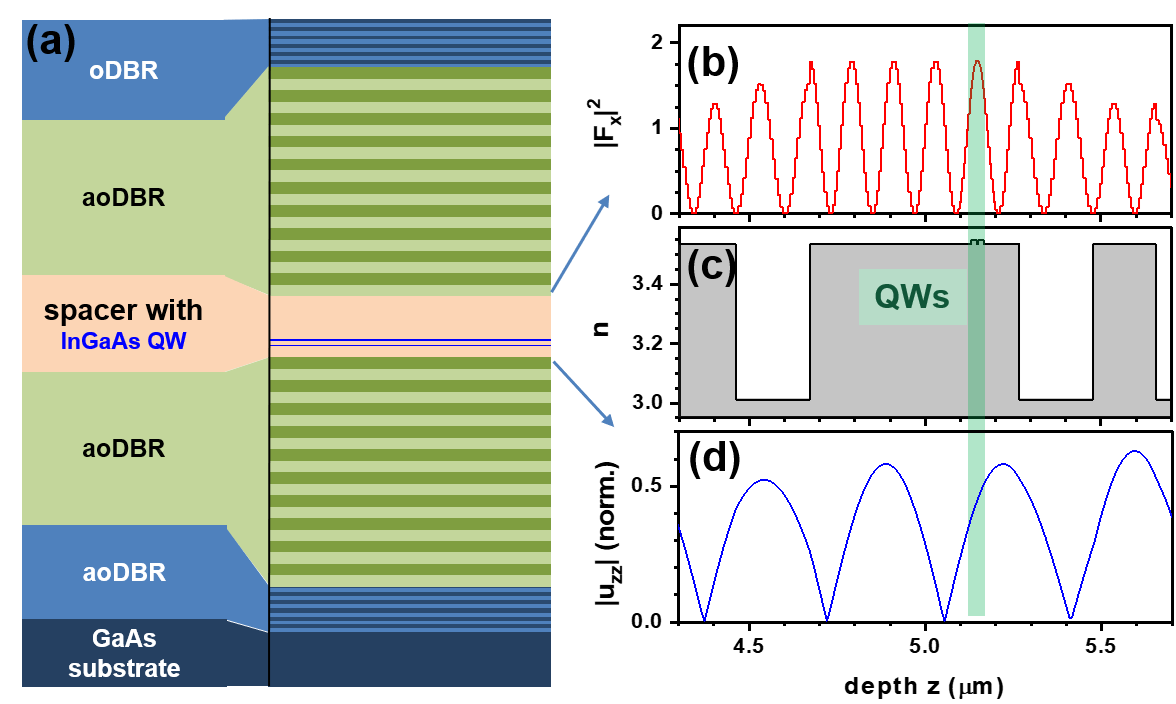}
\caption{
(a) Layer structure of the hybrid microcavity (Sample A). Depth profiles within the spacer region of the hybrid microcavity of the (b) optical field $F_x$ at the photonic resonance energy, where $x$ is the polarization direction on the MC surface, (c) refractive index, $n$,  and (d) normalized acoustic strain field $u_{zz}$ for the acoustic resonance frequency $f_\mathrm{MC} = 6.9$. The green band designates depth of the two quantum wells (QWs).
}
\label{FigS1}
\end{figure}

Fortunately, a very good matching of  $u_{zz}$ and $F_x$ can still be achieved in the hybrid MC of Sample A by slightly displacing the QWs with respect to the anti-nodes of $u_{zz}$. Figure~\ref{FigS1}(b) and (d) shows calculations of the optical ($F_x$) and acoustic ($u_{zz}$) field distributions within the spacer of the MC [cf. Fig.~\ref{FigS1}(a)] carried out using a transfer matrix approach. Figure~\ref{FigS1}(c) shows, for reference, the depth modulation of refractive index $n$ in the same regions indicating the position of the QWs. 
These plots show that the anti-nodes of $u_{zz}$ coincide with the nodes of $F_x$. 
Note that the separation between the QWs is much smaller than the wavelength of both the optical and acoustic fields, so that they can be considered to be subjected to the approximately the same field amplitude. The middle $z$ coordinate of the two QWs is slightly shifted away from the anti-node of $u_{zz}$ to match an anti-node of $F_x$.   The strain field at the QWs is still about {$\eta_s=80$\%}  of its maximum value, so that the small shift only marginally reduces the opto-electronic coupling.

A similar approach was used in Sample B: this sample contains a single (In,Ga)As QW, which was slightly displaced from the anti-nodes of the optical field to ensure a higher coupling to the acoustic field.

\begin{figure}[!thbp]

\includegraphics[width=.7\textwidth, keepaspectratio=true]{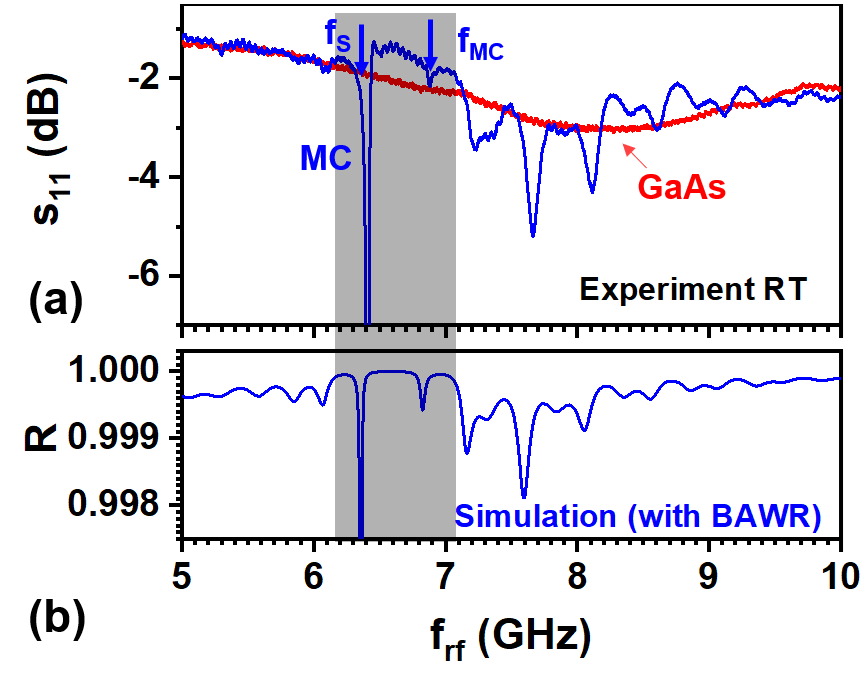}

\caption{
(a) Electrical response of the BAWR on Sample A at 300K. The red curve shows the $s_{11}$ scattering parameter for a BAWR on a bare GaAs substrate, while the blue one gives the response of a device on a MC (Sample A). Both devices have nominally identical ZnO thickness of 260~nm. The gray-shaded area designates the spectral range of acoustic stopband  created by the aoDBR  displayed Fig.~\ref{Fig1}(a) of the main text.  The two modes within the stopband correspond to the MC acoustic mode with $f\textsubscript{MC} = 6.9$~ GHz and the surface mode $f_\textsubscript{S} = 6.4$~GHz confined between the BAWR surface and the upper aoDBR.
(b) Simulated acoustic reflectivity of the MC with a BAWR device on its surface.}
\label{FigS2}
\end{figure}

\section{Frequency response of bulk acoustic resonators}
\label{Frequency response of bulk acoustic resonators}

Figure~\ref{FigS2}(a) compares the frequency response of BAWRs deposited on a plain (001) GaAs substrate (red line) with the one fabricated on top of the MC structure of Sample A. In the former case, the rf-frequency response is dominated by a single broad resonance with a frequency bandwidth of approximately 1~GHz.  On the surface of the MC, the BAWR develops a frequency spectrum characterized by multiple resonances [cf.~Fig.~\ref{FigS2}(a)]. As discussed in the main text, one can identify two acoustic modes -- $f_\textsubscript{S}$ and $f_\textsubscript{MC}$ -- both located within the acoustic stopband of the MC. In this case, the ZnO thickness was larger than the nominal one of 260 nm, resulting in the MC acoustic stopband localized on the shoulder of the BAWR peak [red line in Fig.~\ref{FigS2}(a)]. The experimental results are in  agreement with the transfer-matrix simulations, cf. Fig.~\ref{FigS2}(b) that accurately reproduce all spectral features of acoustic response.

\begin{figure}[!thbp]
\includegraphics[width=.9\textwidth, keepaspectratio=true]{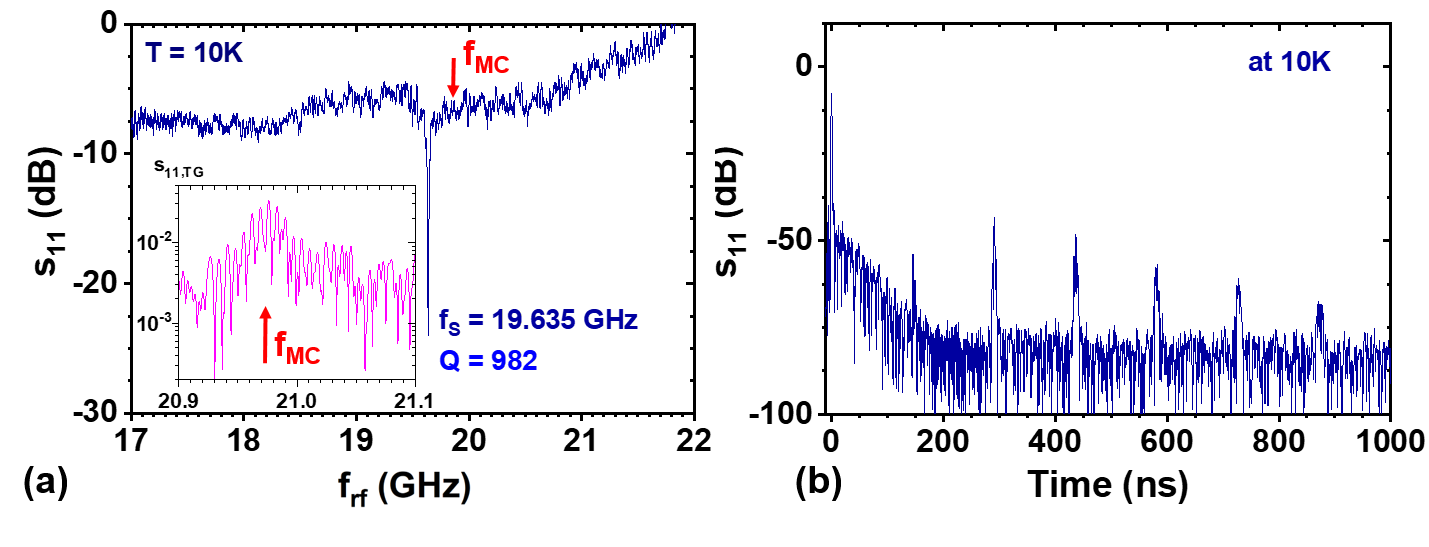}
\caption{
(a) Electrical response of the BAWR on Sample B measured at 10~K, as given by the $s_{11}$ scattering parameter recorded as a function of (a) frequency and (b) time. The inset in (a) displays the acoustic reflection $s_\mathrm{11,TG}$ determined by  Fourier back-transforming the spectrum in (b) within the delay region of the acoustic echoes (i.e., for long delays). The curves were acquired using a BAWR with circular (rather than ring-shaped) contacts with a diameter of 20~$\mu$m. }
\label{FigS3}
\end{figure}

Figures~\ref{FigS3}(a) and \ref{FigS3}(b) compare the electrical response of a BAWR on Sample B in the frequency and time domains, respectively. The measurements were recorded on a BAWR with circular (rather than ring-shaped) electrodes with a diameter of $20~\mu$m: at the high frequencies, the rf-response  of these devices is stronger and  much less noisier than for the ring-shaped ones. In agreement with the results for the 7~GHz devices of Fig.~\ref{FigS2}, the $s_{11}$ spectra shows a sharp dip associated with the MC mode ($f_{MC}$) as well as echoes resulting from multiple reflections of the BAW at the sample boundaries.  The inset in Fig.~\ref{FigS3}(a) displays the acoustic reflection $s_\mathrm{11,TG}$ determined by  Fourier back-transforming the spectrum in (b) within the delay region of the acoustic echoes (i.e., for long delays). The frequency comb within the $f_{MC}$  range contains sharp lines with a line width yielding an acoustic quality factor of $Q_a=6800$.

\subsection*{Quality factor of the acoustic resonances}
\label{Quality factor of the acoustic resonances}

The propagation properties of SHF BAWs  in GaAs (001) substrates at low temperatures has been studied in Ref.~\onlinecite{PVS327}. In these studies, the propagation losses have been expressed in terms of the effective amplitude absorption coefficient $\alpha_\mathrm{eff}$, which includes losses during both propagation and reflection at the sample boarders.
Furthermore, they have shown that the propagation losses at temperatures less than 20~K are mainly associated with losses during reflection at the substrate surfaces. In this case, the effective power absorption fraction per round-trip can be written as  $2r_b=4\alpha_\mathrm{eff} d_\mathrm{Sub}$, where $r_b$ is the power loss fraction per reflection and  $d_\mathrm{Sub}$ denotes the substrate thickness. The Q factor of the comb resonances in  a bare GaAs substrate becomes:

\begin{equation}
    Q_\mathrm{a,Sub} = 2\frac{d_\mathrm{Sub}}{ r_b\lBAW}=\frac{1}{ 2 \alpha_\mathrm{eff}\lBAW }.
    \label{EqSM1}
\end{equation}

\noindent  The previous expression neglects BAW reabsortion by the BAWR. Since $\alpha_\mathrm{eff}$ remains constant within the SHF range, $Q_\mathrm{Sub}$ is expected to increase with the BAW frequency.

The previous analysis can be extended to the MC structures if the acoustic losses at the high-quality MC interfaces can be neglected. In this case, the main effect of the MC is to increase the effective round trip length by $Q_\mathrm{a,MC} d_s$, where $d_s$ denotes the thickness of the MC spacer and $Q_\mathrm{a,MC}$ is the quality factor of the bare MC determined from the reflection coefficient of the DBRs (i.e., by neglecting BAW back-feeding via to reflections at the sample boundaries).  The effective acoustic Q for the comb resonances becomes:

\begin{equation}
    Q_\mathrm{a} = 2\frac{d_\mathrm{Sub}+d_s Q_\mathrm{a,MC}}{ r_b\lBAW}
    =\frac{1}{ 2 \alpha_\mathrm{eff}\lBAW } \left( 1 + \frac{d_s}{d_\mathrm{Sub}} Q_\mathrm{a,MC} \right)
    =Q_\mathrm{a,Sub} \left( 1 + \frac{d_s}{d_\mathrm{Sub}} Q_\mathrm{a,MC} \right)
    \label{EqSM2}
\end{equation}

\noindent Note that for  $Q_\mathrm{a,MC}<<\frac{d_\mathrm{Sub}}{d_s}$, $Q_a$ approaches the substrate quality factor, while for $Q_\mathrm{a,MC}>>\frac{d_\mathrm{Sub}}{d_s}$ there is a considerable enhancement of the quality factor due to acoustic confinement at the MC.

Table~\ref{TS1} compares 
measured quality factors ($Q_a$ (exp)) for samples A and B with the values  estimated  ($Q_a$)  from Eqs.~\ref{EqSM1} and \ref{EqSM2} assuming  $d_s=\lBAW$ and $\alpha_\mathrm{eff} = 8\times 10^{-4}~\mu$m$^{-1}$\cite{PVS327}. The latter yield the quality factors  $Q_\mathrm{a,Sub}$ for the bare substrate listed in the thrid column of the table. In the estimations, we use the values for $Q_\mathrm{a,MC}$  determined from the envelope of the $s_{11}$ response of the BAWR. The higher $Q_a$ for the 20 GHz MC is mainly due to the larger $Q_\mathrm{a,MC}$ arising from the larger number of acoustic DBR stacks. In general, the calculated $Q_a$'s overestimated the measured ones. This discrepancy may be due to the fact that the simple model presented above neglects  losses at the BAWR as well as at   the MC interfaces.

\begin{table}
    \begin{center}
        \begin{tabular}{||c c c c c c||} 
            \hline
    -       & $\fBAW$ (GHz) & $Q_\mathrm{a,Sub}$ & $Q_\mathrm{a,MC}$  & $Q_a$ (calc) & $Q_a$     \\ 
    \hline\hline
    Sample A        & 7     & 3050             & 172               & 4100  &    2800 \\
    \hline
    Sample B        & 20     & 8700             & 982               & 14500  &    6800 \\
    \hline
    \end{tabular}
%
%
\end{center}
    \caption{Measured [$Q_a$] and estimated  [$Q_a$ (calc)] acoustic quality factors  from Eqs.~\ref{EqSM1} and \ref{EqSM2} using $d_s=\lBAW$ and $\alpha_\mathrm{eff} = 8\times 10^{-4}~\mu$m$^{-1}$\cite{PVS327}. $Q_\mathrm{a,MC}$ is the bare quality factor of the acoustic mode determined from the envelope of the $s_{11}$ response of the BAWR. $Q_\mathrm{a,Sub}$ is the quality factor of the bare substrate determined from  $\alpha_\mathrm{eff}$ [cf. Eq.~\ref{EqSM1}].}
    \label{TS1}
\end{table}

\begin{figure}[!thbp]
\includegraphics[width=1\textwidth, keepaspectratio=true]{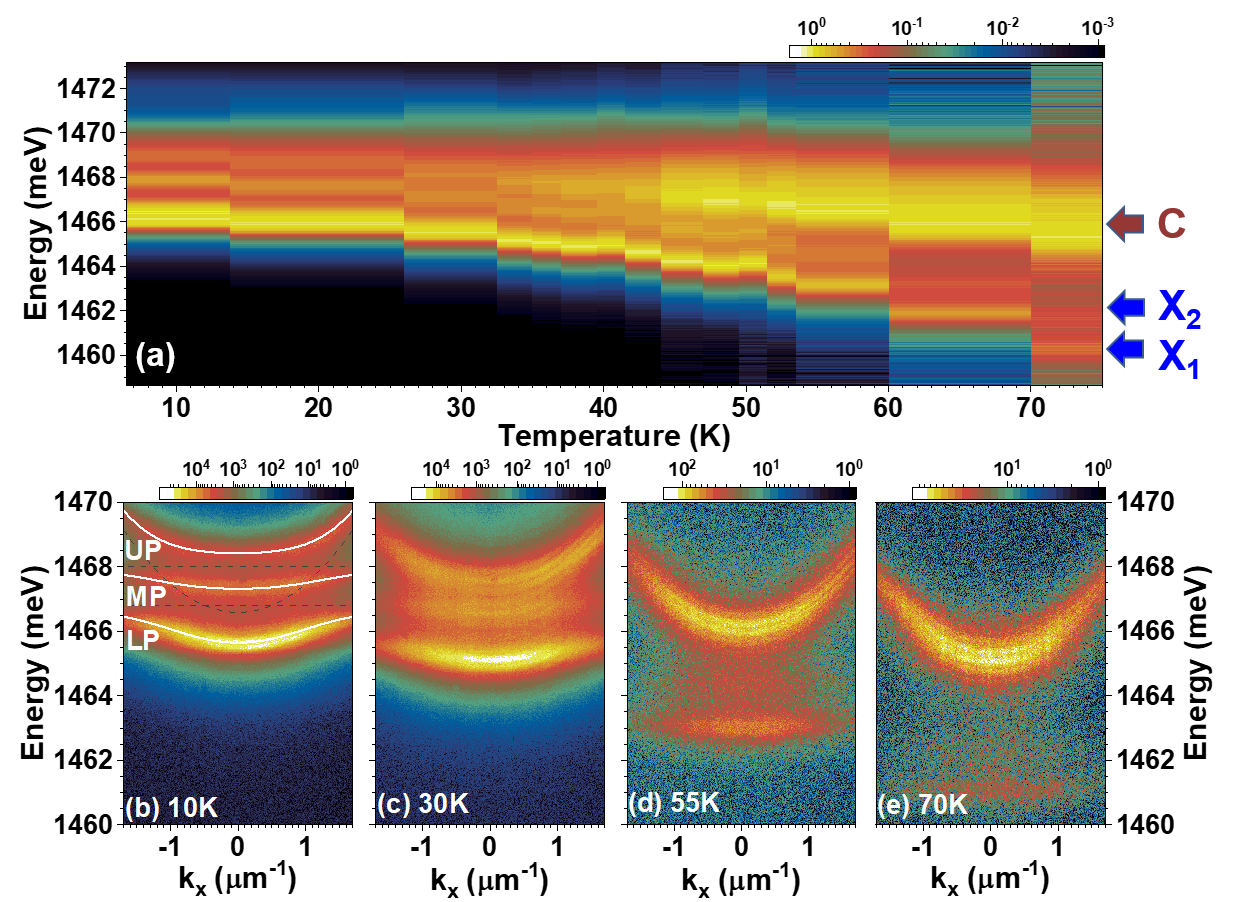}

\caption{
Temperature dependence of the  photoluminescence (PL)  of the hybrid MC (Sample A). (a) Temperature dependence of the PL spectrum. $X_1$ and $X_2$ are excitonic resonances of the coupled InGaAs QWs, which couple to the photonic ($C$) mode of the MC to form the lower (LP), middle (MP) and upper (UP) states at low temperatures. Momentum resolved PL at (b) 10~K, (c) 30~K, (d) 55~K 
and (e) 70~K. Below 50K the system is in the strong-coupling regime. The solid lines 
in (e) are three coupled oscillator fits to the data. The dashed lines are bare energies. The fitted Rabi-splitting energy is $\Omega_\textsubscript{Rabi} = 2 \pm 0.3$~meV.
}
\label{FigS4}
\end{figure}

\section{Temperature dependence of the opto-electronic resonances}
\label{Temperature dependence of the opto-electronic resonances}

The nature of the light-matter coupling in the MCs can be accessed by studying the temperature dependence of the PL. Two InGaAs QWs in Sample A are (unintentionally) tunnel-coupled. This coupling produces excitonic bonding ($X_1$) and anti-bonding ($X_1$) resonances, which are red-shifted with respect to the photonic mode ($C$) at temperature above 50~K. As a consequence, PL spectra recorded at temperatures above 50~K shows  three branches, as illustrated in Fig.~\ref{FigS3}. As the temperature reduces , the excitonic resonances blue-shift and strongly couple to the photonic mode, giving rise to the lower (LP), middle (MP), and upper (UP) polariton branches indicated in Fig.~\ref{FigS3}(a).

The energy dispersion of the $X_1$, $X_2$, and $C$ obtained from angular-resolved PL spectra and displayed Figs.~\ref{FigS3}(b)-(e) for different temperatures gives further evidence for the strong coupling between the excitonic and photonic resonances. The momentum resolved maps of FIG. SM4 were measured by positioning the entrance slit of a single-pass spectrometer in a plane conjugate to the Fourier (back focal) plane of the objective lens. At temperatures above 50~K the photonic resonance shows a strong dispersion, which contrast with the essentially flat dispersion of the excitonic states. This behavior is typical for excitonic resonances in the regime of the weak-coupling to photonic modes. At lower temperatures [cf.~Figs.~\ref{FigS3}(d)-(e)], all resonance lines are dispersive, thus showing that they couple to form polaritons. By fitting the angular-resolved PL map at 10~K to a model of three coupled resonances, we obtain a light-matter Rabi-splitting $\Omega_\mathrm{Rabi} = 2 \pm 0.3$~meV.

\section{Lateral field distribution in ring-shaped BAWR}
\label{Lateral field distribution in ring-shaped BAWR}

The results in the main text prove that the MCs confine BAWs in the direction perpendicular to the surface. Here, we show that the ring geometry of the BAWR displayed in Fig.~\ref{FigS4}(a) confines the acoustic field in the aperture for light access, thus increasing the acoustic field  while providing a favorable geometry for optical access to the active region of the  MCs (i.e., the MC spacer containing the QWs).
 
The investigations of the lateral distribution of the BAW field were carried out by exciting polaritons in Sample A using a 631~nm pulsed laser diode focused onto the center of a BAWR aperture, as shown in Fig.~\ref{FigS4}(a). The MHz-pulsed electrical output of the laser diode controller was used to trigger the rf generator to deliver 1~ns pulse trains of a few uW power to the BAWs. The PL signal was imaged on the slit of a single pass spectrometer, producing spatially resolved PL spectra  across the BAWR aperture.  In order to rule out the heating effect we first measured PL with acoustic ($f_{rf} = 6.9247$~GHz) and laser pulses driven out-of-phase. The corresponding image shows no spectral  modification of the PL collected within the confines of the BAWR aperture [cf.~Fig.~\ref{FigS4}(b)]. 
The studies were carried out at 50~K, where the excitonic lines ($X_1$ and $X_2$ in Sample B) are red-shifted with respect to the phononic mode  ($C$). When the laser pulses are in phase with the rf ones (the in-phase condition), we observe a large spectral change in the detected PL  [cf.~Fig.~\ref{FigS4}(c)]. 
The most pronounced change is the apparent broadening of the spectrum due the modulation of the excitonic resonances with an amplitude $\Delta E$. 
A comparison of the out-of-phase and in-phase spectra at the aperture center shows that the 
main changes arise from the acoustic energy modulation of the excitonic levels with the amplitude $\Delta E$ indicated in the plot [cf.~Fig.~\ref{FigS4}(c)]. 
In addition,  the acoustic modulation induces a decrease of the time-integrated PL intensity since it, in average, energetically shifts the excitonic modes away from the photonic resonance.

The dashed line in Fig.~\ref{FigS4}(c) is a guide-to-the-eye following the maxima of the exciton-related PL intensity, which shows that  that $\Delta E$ slight increases towards the center of the aperture, thus indicating that the ring-geometry concentrates the acoustic field in the center of the aperture. 

\begin{figure*}[!thbp]
\centering\includegraphics[width=\textwidth, keepaspectratio=true]{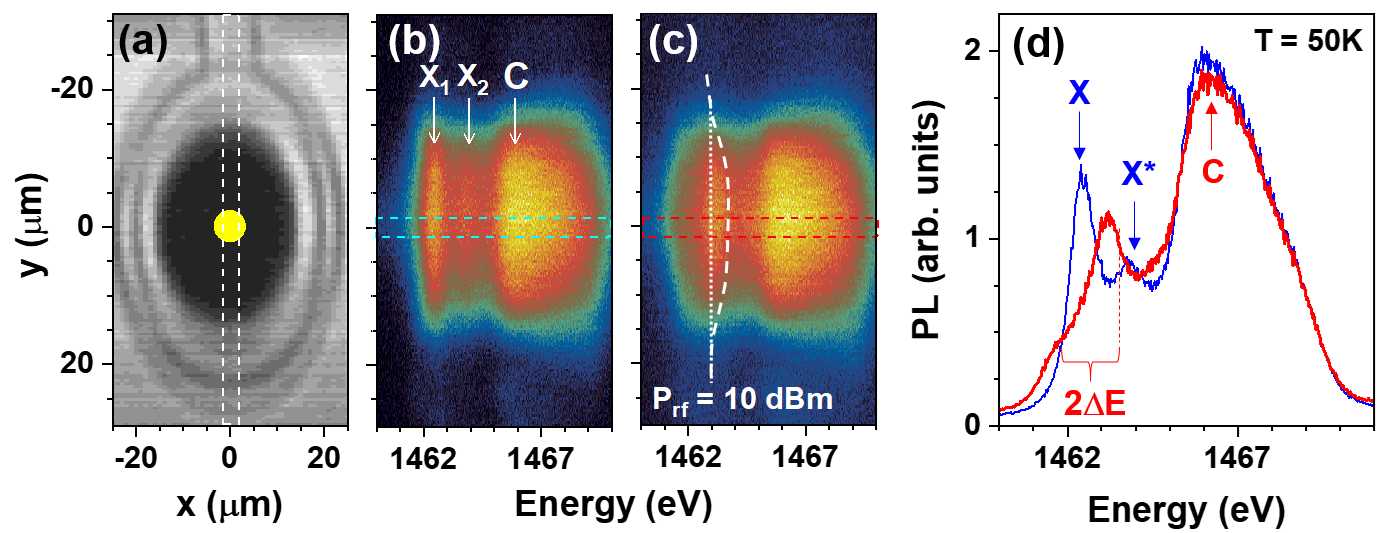}
\caption{Photoluminescence (PL) spectroscopy in the hybrid MC of Sample A with ring-shaped BAWR at 50~K. (a) Optical micrograph of the ring-shaped BAWR indicating the 635~nm laser excitation spot focused at the center of the BAWR aperture. 
(b)-(c) Maps of PL intensity as a function of energy (horizontal axis) and position along the slit (vertical axis) 
without (b) and under BAW excitation of the acoustic MC mode $f_\textsubscript{MC} = 6.9247$~GHz (c). These maps were recorded by collecting the PL emitted within the dashed rectangle in (a). (d) Comparison of PL spectra recorded in the center of the BAWR aperture in the absence (blue) and presence of the BAW (red). The spectra were produced by spatial integration over the regions delimited by dashed lines in (b) and (c), respectively. $\Delta E$ is the energy amplitude modulation of the excitonic resonances. }
\label{FigS5}
\end{figure*}

\begin{figure}[!thbp]

\includegraphics[width=.7\textwidth, keepaspectratio=true]{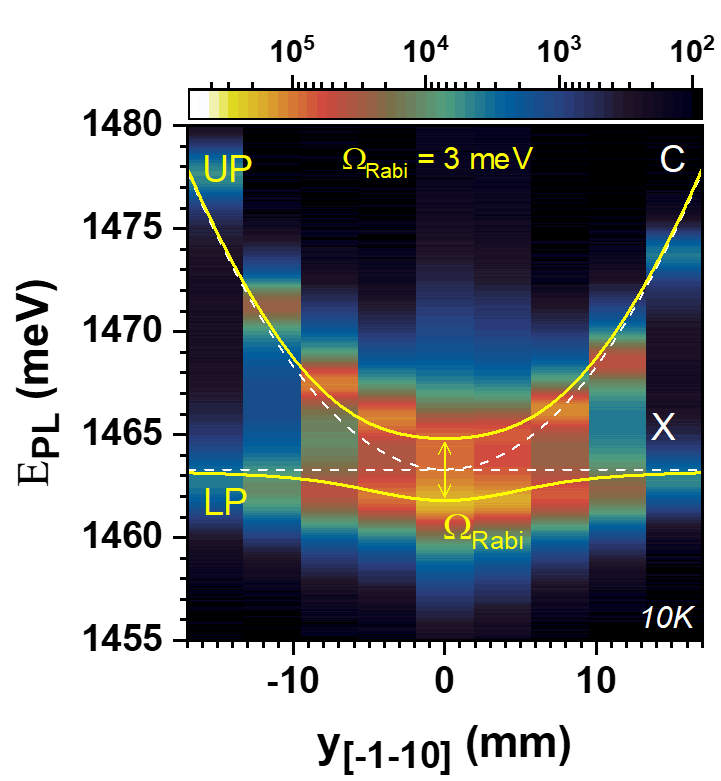}

\caption{
Spatial dispersion of photoluminescence of Sample B at 10 K. The white dashed lines represent calculated spatial dispersion of bare cavity (C) and exciton (X) modes. The solid yellow lines are fits using two coupled oscillators model. The fitted Rabi-splitting energy is $\Omega_\textsubscript{Rabi} = 3 \pm 0.5$~meV.}
\label{FigS6}
\end{figure}

\section{Compensation for thermal effects in the spectroscopic measurements}
\label{Compensation for thermal effects in the spectroscopic measurements}

\begin{figure}[!thbp]

    \includegraphics[width=.7\textwidth, keepaspectratio=true]{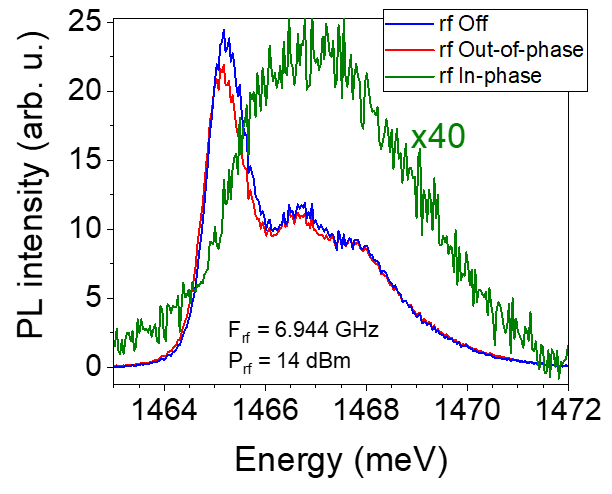}
    
    \caption{
    (a) Comparison of the PL under three different rf-conditions: (i) rf off, (ii) rf out-of-phase and (iii) rf in-phase. The in-phase and out-of-phase are measured at the same resonant rf-frequency 6.944 GHz and the same applied rf-power 14 dBm.}
    \label{FigS7}
    \end{figure}

The rf excitation of the transducers induces sample heating.
The $S_{11}$ coefficient for the device in Fig.~SM2(a) shows that at $f_{MC} ≈ 6.9$~GHz only about 12\% of the applied rf-power is converted into coherent phonons and 88\% are losses, including heating. Since the temperature sensor in the measurement setup is not placed directly on the sample, it cannot detect these changes. Alternatively, PL measurements allow to probe the temperature locally, because the emission intensity and energy of the exciton depend strongly on the temperature. Figure~SM4(a) shows that we can detect temperature changes with an accuracy of 5~K. 

In order to discriminate possible thermal effects arising from sample heating from the ones due to the coherent modulation by the BAWs, the acoustic PL measurements were carried out in the following way~\cite{PVS282}. The rf-generator and the pulsed semiconductor laser-diode exciting the PL were triggered by a train of low frequency square pulses (with frequencies between 1 to 10 kHz). To rule out the effect of temperature, the measurements were carried out under two configurations for the relative phases between the pulse trains: (i) the in-phase condition, where the rf and optical excitation are present at the same time; (ii) out-of-phase – no temporal overlap between the rf and optical pulses. Due to the high pulse frequency, the sample temperature is the same for the two situations. Thus, the out-of-phase contains information about thermal phonons (heat). This method allowed us to rule out the temperature contribution to the PL modulation shown in Fig.~\ref{Fig4} on the main text.

Figure~\ref{FigS7}  compares PL spectra of a device with rf off, rf in-phase and rf out-of-phase (see Methods). The in-phase and out-of-phase are measured at the same resonant rf-frequency of 6.944~GHz and the same applied rf-power of 14~dBm. Thus, the out-of-phase spectrum contains information about thermal phonons (heat). Comparison of the rf off and rf out-of-phase shows that the effect of the heating is negligible compared to the strain-induced modulation. For completeness, the green curve shows a PL spectrum recorded for the rf in-phase condition, which proves that the changes induced by the acoustic modulation are completely different from the purely thermal ones.
\end{document}